\documentclass[paper]{aa}

\usepackage{graphicx}
\usepackage{aalongtable,lscape}
\def\ls{{_<\atop^{\sim}}}

\def\cgs{ ${\rm erg~cm}^{-2}~{\rm s}^{-1}$ } 

\begin{document}
%%%%%%%%%%%%%%

\title{The Swift Serendipitous Survey in deep XRT GRB fields
(SwiftFT)\thanks{The survey's acronym remembers the satellite Swift and Francesca Tamburelli (FT), who contributed in a crucial way to
the development of the {\it Swift}-XRT data reduction software. We dedicate this work to her memory.}\\
I. The X-ray catalog and number counts}

\author{S. Puccetti\inst{1},  M. Capalbi\inst{1}, P. Giommi\inst{1}, M. Perri\inst{1}, G. Stratta \inst{1}, L. Angelini\inst{2}, D. N. Burrows\inst{3}, S. Campana\inst{4}, G. Chincarini\inst{4,5}, G. Cusumano\inst{6}, N. Gehrels\inst{2},  A. Moretti\inst{4}, J. Nousek\inst{3},  J. P. Osborne\inst{7}, G. Tagliaferri\inst{4}}

\institute { $^1$ASI  Science Data Center, via Galileo Galilei, 00044, Frascati,  Italy.\\
$^2$NASA/Goddard Space Flight Center, Greenbelt, MD, USA.\\
$^3$Department of Astronomy and Astrophysics, Pennsylvania State University, 525 Davey Lab, University Park, PA 16802, USA.\\
$^4$INAF, Osservatorio Astronomico di Brera, via E. Bianchi 46, I-23807 Merate (LC), Italy.\\
$^5$Universita` degli Studi di Milano-Bicocca, Dipartimento di Fisica, Piazza delle Scienze 3, I-20126 Milano, Italy.\\
$^6$INAF, Istituto di Astrofisica Spaziale e Fisica Cosmica di Palermo, via U. La Malfa 153, I-90146 Palermo, Italy.\\\
$^7$Department of Physics \& Astronomy, University of Leicester, Leicester LE1 7RH, UK.\\
\email{puccetti@asdc.asi.it}
}

\date{21 January 2011}

\abstract
{}
{An accurate census of the active galactic nuclei (AGN) is a key step
in investigating the nature of the correlation between the
growth and evolution of super massive black holes and galaxy
evolution. X-ray surveys provide one of the
most efficient ways of selecting AGN.}
{We searched for X-ray serendipitous sources in over
370 Swift-XRT fields centered on gamma ray bursts detected between
2004 and 2008 and observed with total exposures ranging from 10 ks to
over 1 Ms. This defines the Swift Serendipitous Survey in deep XRT
GRB fields, which is quite broad compared to existing surveys ($\sim$33
square degrees) and medium depth, with a faintest flux limit of
7.2$\times10^{-16}$ erg cm$^{-2}$ s$^{-1}$ in the 0.5 to 2 keV energy
range (4.8$\times10^{-15}$ erg cm$^{-2}$ s$^{-1}$ at 50\%
completeness). The survey has a high degree of uniformity thanks to
the stable point spread function and small vignetting correction
factors of the XRT, moreover is completely random on the sky as GRBs explode
in totally unrelated parts of the sky.}
{In this paper we present the sample and the X-ray number counts of
the high Galactic-latitude sample, estimated with high statistics over
a wide flux range (i.e., 7.2$\times10^{-16}\div\sim5\times10^{-13}$
erg cm$^{-2}$ s$^{-1}$ in the 0.5-2 keV band and
3.4$\times10^{-15}\div\sim6\times10^{-13}$ erg cm$^{-2}$ s$^{-1}$ in
the 2-10 keV band). We detect 9387 point-like sources with a detection
Poisson probability threshold of $\le2\times10^{-5}$, in at
least one of the three energy bands considered (i.e. 0.3-3 keV, 2-10 keV,
and 0.3-10 keV), for the total sample, while 7071 point-like sources
are found at high Galactic-latitudes (i.e. $\mid$b$\mid\ge$20
deg.). The large number of detected sources resulting from the
combination of large area and deep flux limits make this survey a new
important tool for investigating the evolution of AGN.  In particular,
the large area permits finding rare high-luminosity objects like QSO2,
which are poorly sampled by other surveys, adding precious information
for the luminosity function bright end. The high Galactic-latitude
logN-logS relation is well determined over all the flux coverage, and
it is nicely consistent with previous results at 1$\sigma$ confidence
level.  By the hard X-ray color analysis, we find that the Swift
Serendipitous Survey in deep XRT GRB fields samples relatively
unobscured and mildly obscured AGN, with a fraction of obscured
sources of $\sim37\%$ ($\sim15\%$) in the 2-10 (0.3-3 keV) band.}
{}
\keywords{X-ray: general, Surveys, Catalogs, Galaxies: active.}

\authorrunning {Puccetti et al.}
\titlerunning {The {\it Swift} Serendipitous Survey in deep XRT GRB fields}

\maketitle

%%%%%%%

\section{Introduction}

The ``feedback'' between the super-massive black holes (SMBH), which
fuel the active galactic nuclei (AGN), and the star formation in the
host galaxy tightly links the formation and evolution of AGN and
galaxies. Therefore, a complete knowledge of the evolution and the
phenomena in the AGN is a key topic in cosmology. A good
way to complete the census of AGN is to use X-ray surveys, because
these efficiently select AGN of many varieties at higher sky surface
densities than at other wavelengths. The better capability of finding AGN
in X-rays rests on three main causes: 1) X-rays directly trace
accretion onto SMBH, providing a selection criterion that is less
biased than the AGN optical emission line criterion; 2) AGN are the
dominant population in the X-rays, because they are most ($\sim$ 80\%) of the
X-ray sources; 3) X-rays in the medium-hard band (i.e. $\sim$ 0.5-10
keV band, that is the energy coverage of {\it Chandra} and {\it
XMM-Newton}) are able to detect unobscured and moderately obscured
AGN (i.e. N$_H$$\ls$ a few 10$^{23}$ cm$^{-2}$, Compton thin AGN).

\begin{figure}
\begin{center}
\includegraphics[angle=0,height=8truecm]{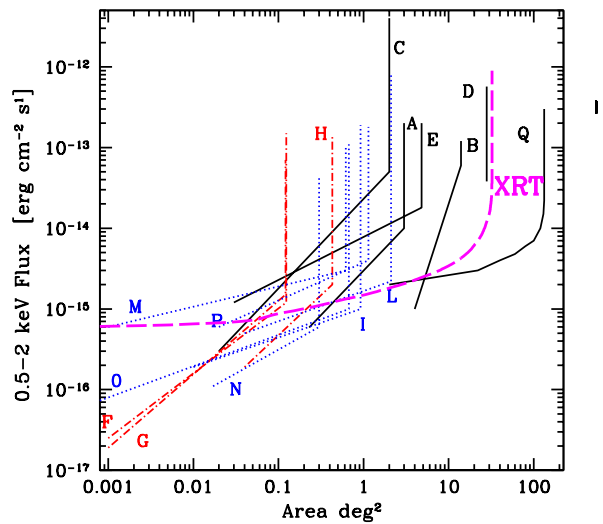}
\caption{The flux limit in the 0.5-2 keV band vs. the area coverage
for various surveys. Magenta long dashed lines are the total XRT Deep
Serendipitous Survey; black solid lines are medium large not
contiguous surveys: A: H2XMM, B: CHAMP, C: SEXSI, D: XMM-BBS, E: AXIS
, Q: twoXMM ($|$b$|>$20); red dot short-dashed lines are the smallest
and very deep surveys: F: CDFN, G: CDFS, H: LockmanHole; blue dotted
lines are shallow contiguous survey I: CCOSMOS, L: XMMCOSMOS, M:
ELAIS-S1, N: ECDFS, O: AEGISX, P: SXDS (for references see Table 1).}
\label{surveys}
\end{center}
\end{figure}

{\it Chandra} and {\it XMM-Newton} performed several deep pencil beam
surveys and shallow wide contiguous surveys (see
Fig. \ref{surveys}). Deep pencil beam surveys (see Table \ref{decsu})
are fundamental in studying the population of faintest X-ray sources,
especially the emerging new population of ``normal'' galaxies (Brandt
\& Hasinger 2005); however, because they sample very small sky
regions, they are strongly affected by cosmic variance. Wide shallow
contiguous surveys (see Table \ref{decsu}) are complementary to deep
pencil beam surveys, since they are less affected by cosmic variance,
by covering a much larger area of the sky. Nevertheless they only
reach relatively high fluxes, losing a large fraction of faint AGN.

The gap between deep pencil beam surveys and the wide contiguous
shallow surveys is filled by the very large, non-contiguous,
medium-depth surveys. This type of survey, based on the large archival
data available from {\it Chandra} and {\it XMM-Newton} satellites (see
Table \ref{decsu}), covers very large sky area, thus finding rare
objects, like the highest luminosity, obscured AGN, QSO2 (see
e.g. HELLAS2XMM, Fiore et al. 2003). An additional fundamental
advantage of this type of survey is the ability to investigate field
to field variations of the X-ray source density, which may trace
filaments and voids in the underlying large-scale structure.

\begin{table*}
\footnotesize
\caption{Main properties of the most famous X-ray surveys ( see Fig. 1).}
\begin{center}
\begin{tabular}{lccccc}

\hline
Label$^a$ & Name & Area &  Flux limit & Band$^b$ & Reference\\
&& deg.$^2$ &  erg cm $^{-2}$ s$^{-1}$ & keV & \\
\hline

\multicolumn{6}{c}{Examples of deep pencil beam surveys}\\
\hline
G & CDFS & $\sim$0.1 & 1.9$\times$10$^{-17}$ & 0.5-2 & Giacconi et al. 2001, Luo et al. 2008\\
F & CDFN  & $\sim$0.1 & 2.5$\times$10$^{-17}$ & 0.5-2 & Brandt et al. 2001, Alexander et al. 2003\\
H & XMM-Newton Lockman Hole & $\sim$0.43 & $\sim$ 1.9$\times$10$^{-16}$ & 0.5-2 & Worsley et al. 2004, Brunner et al. 2008\\

\hline
\multicolumn{6}{c}{Examples of wide shallow contiguous surveys}\\
\hline
N & E-CDF-S & $\sim$0.3 & 1.1$\times$10$^{-16}$ & 0.5-2  & Lehmer et al. 2005\\
M & ELAIS-S1 & $\sim$0.6 & 5.5$\times$10$^{-16}$ &  0.5-2  &  Puccetti et al. 2006\\
L & XMM-COSMOS & $\sim$2 &  1.7$\times$10$^{-15}$ & 0.5-2 &Hasinger et al. 2007, Cappelluti et al. 2007, 2009\\
I & C-COSMOS & $\sim$1 & 1.9$\times$10$^{-16}$ &  0.5-2 & Elvis et al. 2009\\
O & AEGIS-X  & 0.67 &  5$\times$10$^{-17}$  & 0.5-2 &Laird et al. 2009\\

\hline
\multicolumn{6}{c}{Examples of surveys based on serendipitous sources in archival data}\\
\hline
A & Hellas2XMM & 3 & 5.9$\times$10$^{-16}$ & 0.5-2  & Baldi et al. 2002\\
C & SEXSI & $\sim$2 &  5$\times$10$^{-16}$ &  2-10  & Harrison et al. 2003\\
D & XMM-BSS & 28.1 & 7$\times$10$^{-14}$ & 0.5-4.5  & Della Ceca et al. 2004\\
E & AXIS & 4.8 & $\sim$2$\times$10$^{-15}$ & 2-10  & Carrera et al. 2007 \\
P & SXDS & 1.14 & 6$\times$10$^{-16}$ & 0.5-2  & Ueda et al. 2008\\
B & CHAMP & $\sim$10 & $\sim$10$^{-15}$ & 0.5-2  & Kim et al. 2007\\
Q & TwoXMM ($\mid$b$\mid$$\ge20$) & $\sim$132.3 & $\sim$2$\times$10$^{-15}$ & 0.5-2  & Mateos et al. 2008\\
\\
\hline

\end{tabular}

\end{center}

$^{a}$ Label refers to Fig. 1;

$^{b}$ The flux limit is related to this energy band.

\label{decsu}
\end{table*} 

We built a new large medium-depth X-ray survey searching for
serendipitous sources in images taken by the {\it Swift} (Gehrels et
al. 2004) X-ray telescope {\it XRT} (Burrows et al. 2005) centered on
gamma-ray bursts (GRBs).  The {\it Swift} Serendipitous Survey in deep
XRT GRB fields (SwiftFT\footnote{The survey's acronym remembers
the satellite Swift and Francesca Tamburelli (FT), who contributed in
a crucial way to the development of the {\it Swift}-XRT data reduction
software. We dedicate this work to her memory.}) presents significant
advantages compared with present large area X-ray surveys.  First,
{\it Swift} is a mission devoted to discovering GRBs and following
their afterglows, which in X-rays last typically several days after
the burst, so the same sky region can be observed for very long
exposure (as long as $\sim$1.17 Ms in the case of GRB060729). This,
together with the very low and stable background of the XRT camera
($\sim$0.0002 counts sec$^{-1}$ arcmin$^{-2}$ in the 0.3-3 keV band)
permits us to have flux limit of $\sim$7.2$\times$10$^{-16}$ erg
cm$^{-2}$s$^{-1}$ in the 0.5-2 keV (50\% completeness flux limit
of 4.8$\times10^{-15}$ erg cm$^{-2}$ s$^{-1}$, for conversion from
rate to flux see Sec. 4.3), one of the deepest flux limits of any
large area survey. Second, the XRT point spread function and
vignetting factor, being approximately independent of the distance
from the aim point of the observation (i.e. off-axis angle),
secure a uniform sky coverage. This uniform sensitivity provides the
largest area coverage at the lowest flux limits (see
Fig. \ref{surveys}). Third, since GRBs explode randomly on the sky,
with an isotropic distribution (Briggs et al. 1996), the SwiftFT does
not suffer any bias toward previously known bright X-ray sources, as
the large serendipitous surveys based on X-ray archival data, like
{\it Einstein}, {\it ROSAT}, {\it Chandra} and {\it XMM-Newton} data
(see also Moretti et al. 2009). Specifically, a correlation length of
1-10 Mpc corresponds to $\sim$2-20 arcmin at the mean redshift of the
{\it Swift} GRBs (i.e. z$=2.1\pm1.5$, in a cosmological model
($\Omega_M$, $\Omega_\lambda$)=(0.3,0.7)) and to $\sim$10-100 arcmin
at the typical redshift of known X-ray targets (i.e. z$\le$0.1). This
implies that in the case of GRBs, the detection of serendipitous
sources, that might be associated with large scale structure around
the target, is less probable (see e.g. D'Elia et al. 2004 and
references therein).

In this paper we report on the strategy, design and execution of the
SwiftFT: in Sect. 2 we give an overview of the survey and briefly
present the analyzed observations, in Sec. 3 and 4 we describe the
data reduction, detection method and source characterization
procedure, respectively. In Sect. 5 we show the catalog of the
point-like X-ray sources. For the high Galactic-latitude fields
(i.e. $\mid$b$\mid$$\ge$20 deg), we present the survey sensitivity,
the X-ray number counts (i.e. LogN-LogS) and the hardness ratio
analysis in Sec. 6. Finally Sec. 7 shows our conclusion.

\section{ The {\it Swift} Serendipitous Survey in deep XRT GRB fields}

\begin{figure}
\begin{center}
\includegraphics[angle=270,width=9.6truecm]{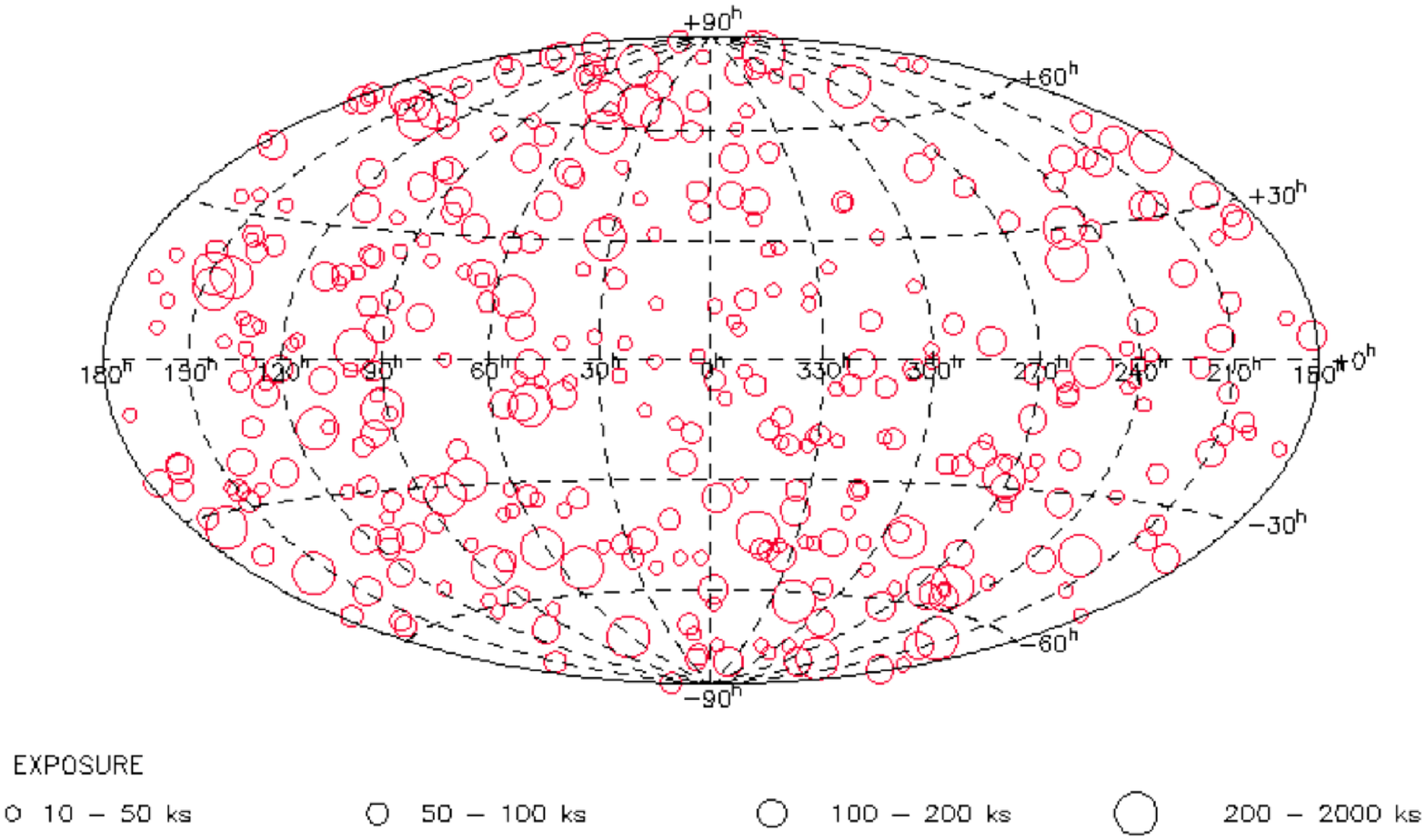}
\caption{Aitoff Projection in Galactic coordinates of the 374
SWIFT-XRT fields analyzed so far.
The dot sizes are proportional to the total field exposure time.}
\label{skymap}
\end{center}
\end{figure}

\begin{figure}
\includegraphics[width=7truecm]{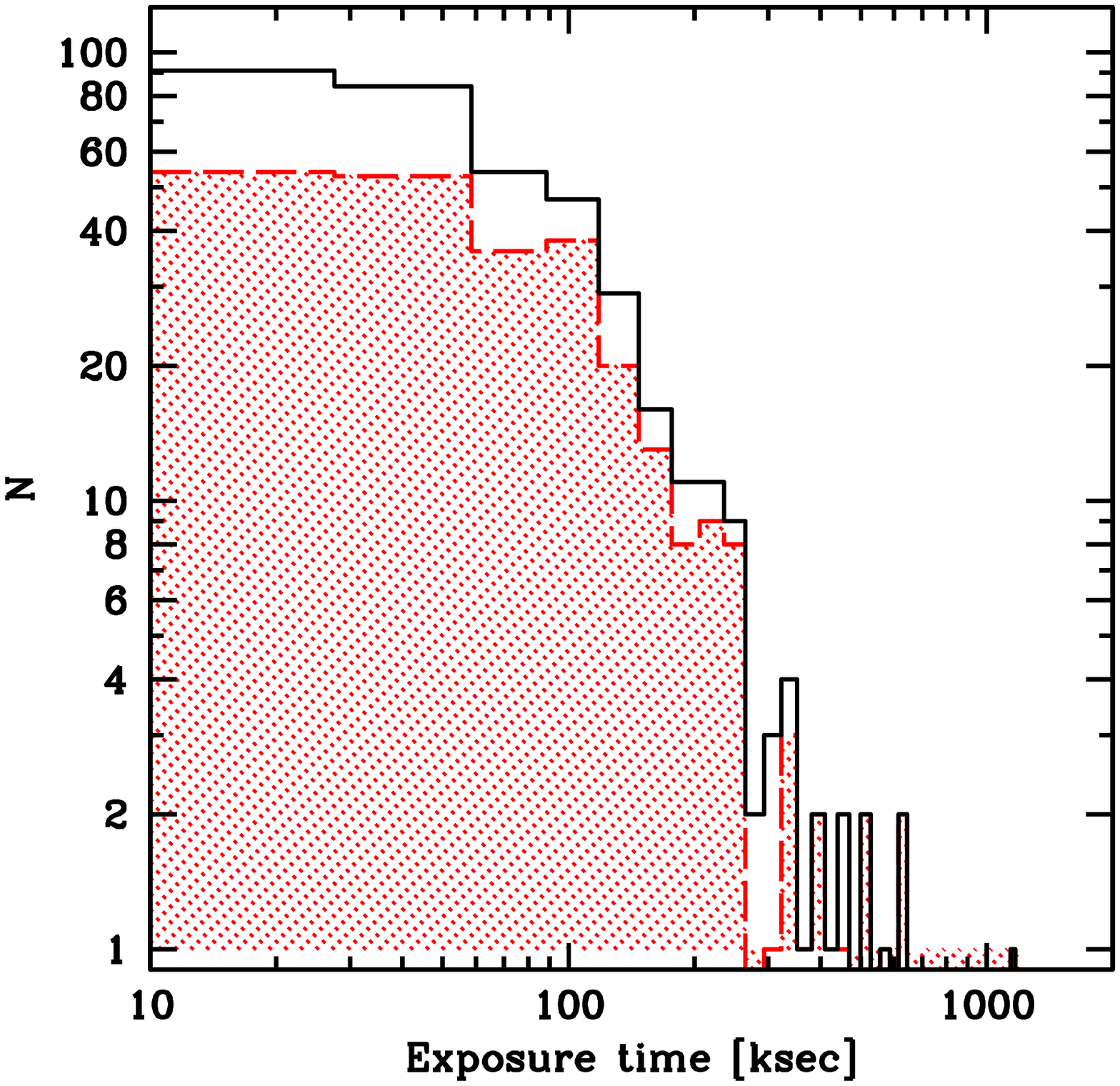}
\includegraphics[width=7truecm]{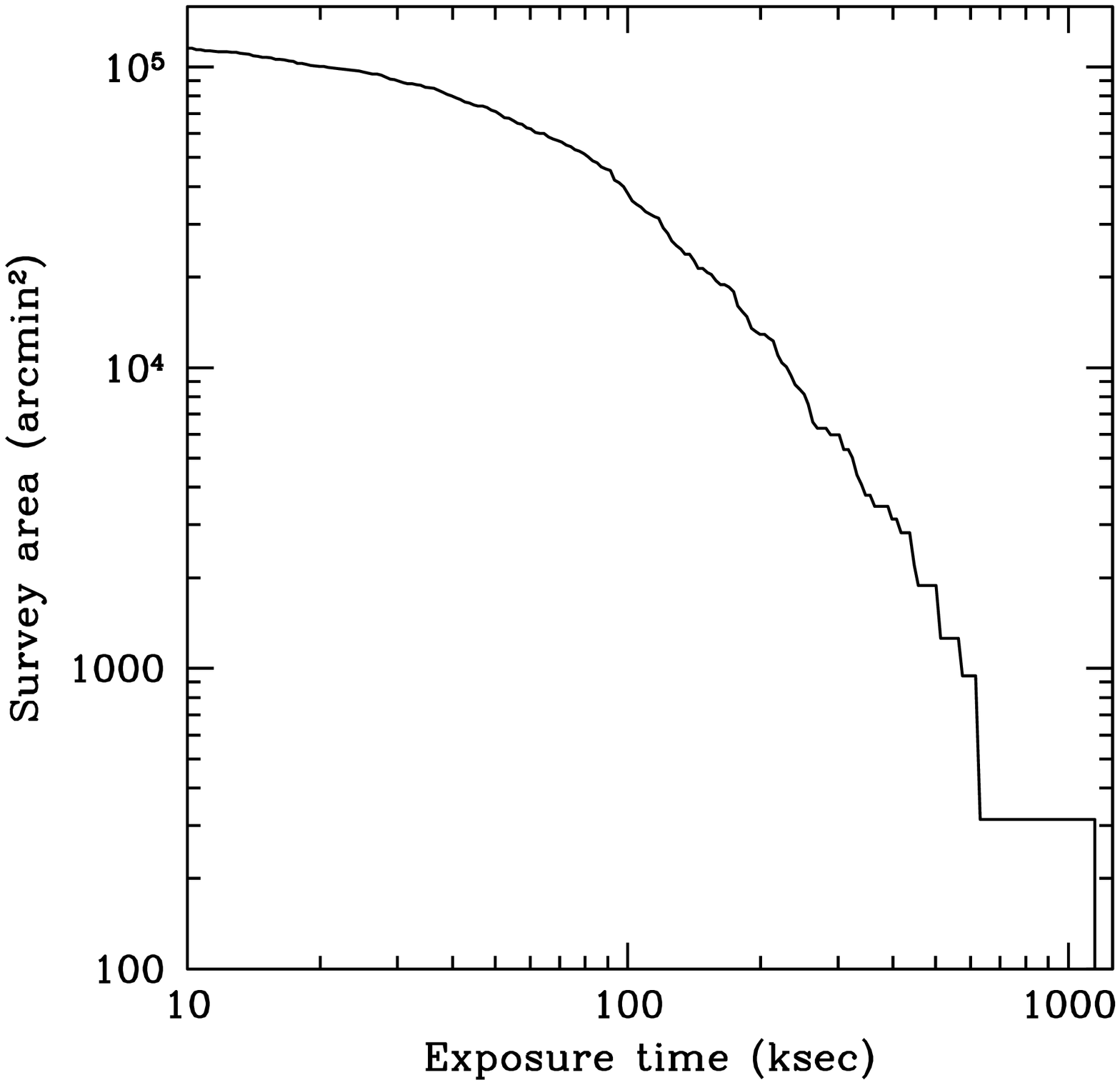}
\caption{ {\it Top panel}: the distribution of the field's exposure times in ks, for the
total sample (black solid histogram) and the HGL sample (red dashed
histogram).{\it Bottom panel}: the survey areas vs the effective exposure time.
}
\label{expoa}
\end{figure}

The {\it Swift} mission (Gerhels et al. 2004) is a multi-wavelength
observatory dedicated to GRB astronomy. {\it Swift}'s Burst Alert
Telescope (BAT) searches the sky for new GRBs, and, upon discovering
one, triggers an autonomous spacecraft slew to bring the burst into
the X-ray Telescope (XRT) and Ultraviolet/Optical Telescope (UVOT)
fields of view. XRT and UVOT follow the GRB afterglow while it remains
detectable, usually for several days. This is achieved by performing
several separate observations of each GRB. By stacking individual
exposures it is possible to build a large sample of deep X-ray
images. To this purpose, we considered all GRBs observed by {\it
Swift} from January 2005 to December 2008, with a total exposure time
in the XRT longer than 10 ks. We also analyzed the XRT 0.5 Ms
observations of the Chandra Deep Field-South (CDFS) sky region.  We
call this set of observations the {\it Swift} Serendipitous Survey in
deep XRT GRB fields (SwiftFT). As GRBs explode at random positions in
the sky the pointing positions of the 374 fields selected in this way
are completely random as shown in Fig. \ref{skymap}. The total
exposure time is 36.8 Ms, with $\sim$32\% of the fields having more
than 100 ks exposure time, and $\sim$28\% with exposure time in the
range 50-100 ks (see top panel of Fig. \ref{expoa}).  The SwiftFT
covers a total area of $\sim$32.55 square degrees; the bottom panel of
Fig. \ref{expoa} shows the exposure time versus the survey area. A
complete list of the fields is available on-line at this address
http://www.asdc.asi.it/xrtgrbdeep\_cat/logXRTFIELDS.pdf. This table
for each field gives the field name, the R.A., the Dec., the
start-DATE, the end-DATE and the total exposure time.\\ In this paper
we concentrate on extragalactic X-ray sources so we consider in detail
the 254 fields at high Galactic-latitudes ($\mid$b$\mid\ge$20, HGL
catalog hereinafter), which cover a total area of $\sim$22.15 square
degrees and have a total exposure time of 27.62 Ms (see
Fig. \ref{skymap} and \ref{expoa}).

\section{XRT data reduction}

The XRT data were processed using the XRTDAS software (Capalbi et
al. 2005) developed at the ASI Science Data Center and included in the
HEAsoft 6.4 package distributed by HEASARC.  For each field of the
sample, calibrated and cleaned Photon Counting (PC) mode event files
were produced with the {\it xrtpipeline} task.
In addition to the screening criteria used by the standard pipeline
processing, a further, more restrictive, screening was applied to
the data, in order to improve the signal to noise ratio of the faintest,
background dominated, serendipitous sources.

Therefore we selected only time intervals with CCD temperature less
than $-50$ $\deg$C (instead of the standard limit of $-47$~$\deg$C)
since the contamination by dark current and hot pixels, which increase
the low energy background, is strongly temperature dependent.
Moreover, to exclude the background due to residual bright earth
contamination, we monitored the count rate in four regions of
70$\times$350 physical pixels, located along the four sides of the
CCD. Then, through the {\it xselect} package, we excluded time
intervals when the count rate was greater than 40 counts/s.  This
procedure allowed us to eliminate background spikes, due to scattered
optical light, that usually occur towards the end of each orbit when
the angle between the pointing direction of the satellite and the
day-night terminator (i.e. bright earth angle, BR\_EARTH) is low.

We performed the on-ground time dependent bias adjustment choosing, in
each time interval, a single bias value using the entire CCD window
and we applied this value to all the events collected during the time
interval. Finally we note that multiple observations of the same field
may differ somewhat in aim point and roll angle. In order to have a
uniform exposure, we restricted our analysis to a circular area of 10
arcmin radius, centered in the median of the individual aim
points. The observations of each field were processed providing an
input to the {\it xrtpipeline} of a fixed pointing direction chosen as
the median of the different pointings on the same target. The cleaned
event files obtained with this procedure were merged using xselect.
In some of the deepest images of our sample ($> 200$ ks) we found
evidence of several hot pixels along the detector column DETX$=$295;
therefore we excluded this column from our analysis.

\begin{figure}
\begin{center}
\includegraphics[angle=0,height=7truecm]{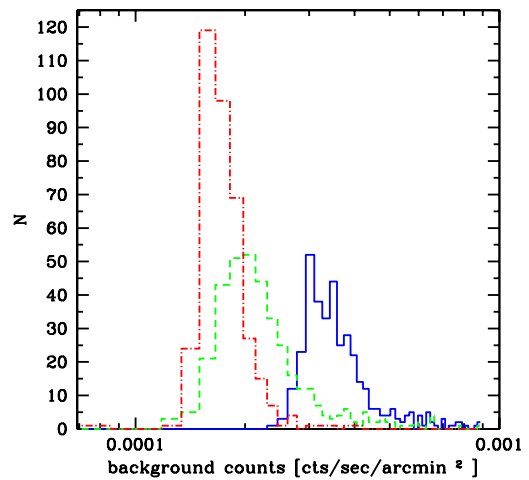}
\caption{The distribution of the mean background counts/sec/arcmin$^2$
for the 374 XRT fields, in the F band (blue solid histogram), S band
(green dashed histogram), and H band (red dot-dashed histogram). }
\label{backf}
\end{center}
\end{figure}

As for the event files, we produced exposure maps of the
individual observations, providing as input to the {\it xrtexpomap} a
fixed pointing direction equal to the median of the pointings on the same target. 
The corresponding total exposure maps were generated by summing the exposure maps of the individual
observations with {\it XIMAGE}. 
We produced exposure maps at three energies: 1.0 keV, 4.5 keV, and 1.5 keV. These correspond to
the mean values for a power-law spectrum of photon index
$\Gamma$$=1.8$ (see Sec 4.3) weighted by the XRT efficiency over the
three energy ranges: 0.3-3 keV (soft band S), 2-10 keV (hard
band H), 0.3-10 keV (full band F) considered.\\
For each field we also produced a background map, using {\it XIMAGE}
by eliminating the detected sources and calculating the mean
background in box cells of size 32$\times$32 pixels. Fig. \ref{backf}
shows the distribution of the mean background counts/s/arcmin$^2$ in
the three energy bands: S, H and F. The median values of background
and their interquartile range are 0.22$\pm$0.04
counts/ks/arcmin$^2$, 0.17$\pm$0.01 counts/ks/arcmin$^2$ and
0.35$\pm$0.05 counts/ks/arcmin$^2$ for the S, H and F band,
respectively. These median values correspond to a level of less than 1
count in the S, H, and F band, over a typical source detection cell
(see Sec. 4.1) and exposure of 100 ks. The low background is
important for the detection of the faintest sources.

\section{Data analysis}

\subsection{Source Detection}

The X-ray point source catalog was produced by the detection algorithm
{\it detect}, a tool of the {\it XIMAGE} PACKAGE version 4.4.1
\footnote{
http://heasarc.gsfc.nasa.gov/docs/xanadu/ximage/ximage.html}. {\it
Detect} locates the point sources using a sliding-cell method. The
average background intensity is estimated in several small square
boxes uniformly located within the image. The position and intensity
of each detected source are calculated in a box whose size maximizes
the signal-to-noise ratio. The net counts are corrected for dead times
and vignetting using the input exposure maps, and for the fraction of
source counts that fall outside the box where the net counts are
estimated, using the PSF calibration. Count rate statistical and
systematic uncertainties, are added quadratically. We set {\it detect}
to work in {\it bright} mode, that is recommended for crowded fields
and fields containing bright sources, since it does not merge the
excess before optimizing the box centroid (see the {\it XIMAGE}
help). We tested {\it detect} on a sample of fields, with deep and
medium deep exposure times, to determine the other detection
parameters, that are the most suitable for our survey. To this end we
also compared the results of {\it detect} applied to the field
GRB070125, with the results of the {\it ewave} detection algorithm of
{\it CIAO}\footnote{ http://cxc.harvard.edu/ciao/ } applied to the
{\it Chandra} observations of the same field. We found that background
is well evaluated for all exposure times, using a box size of
32$\times$32 original detector pixels, and that the optimized size of
the search cell, that minimizes source confusion, is 4$\times$4
original detector pixels. We also set the signal-to-noise acceptance
threshold to 2.5. We produced a catalog using a Poisson probability
threshold of 4$\times10^{-4}$. Here we present only a more
conservative catalog cut to a probability threshold of
2$\times10^{-5}$, to minimize the number of spurious sources. This
probability corresponds to about 0.24 spurious sources for each field
(see Sec. 4.2)\\ We applied {\it detect} on the XRT image using the
original pixel size, and in the three energy bands: F, S and H (see
Sec. 3). For each field we detected only sources in a circular area of
10 arcmin radius centered in the median of the individual aim points
(see Sect. 3). We find that a straight application of {\it detect} on
those images to which the spatial filter was applied leads to an
incorrect estimate of the count rates from the sources near the edges
of the circular area; this is a consequence of the inaccurate PSF
correction and a poorly estimated background at the image edges. To
overcome this difficulty, we applied a two step spatial filter. We
first ran {\it detect} on the images to which the spatial filter was
applied, to select only a circular area of 10.5 arcmin radius centered
at the median of the individual aim points. Then, we applied a second
spatial filter to the catalog, accepting only sources whose distance
from the image center is equal to, or less than, 10 arcmin.\\ This
catalog was cleaned from obvious spurious sources, like detection on
the wings of the PSF or near the edges of the {\it XRT} CCD, spurious
fluctuations on extended sources etc., through visual inspection of
the {\it XRT} images in the three energy bands. We eliminated the GRBs
by matching the catalog with the GRB positions by Evans et
al. 2009. Moreover we also eliminated extended sources from the final
point-like catalog, because {\it detect} is not optimized to detect
this type of sources, not being calibrated to correct for the
background and PSF loss in case of extended sources (a detailed
catalog of the extended sources will be presented in a forthcoming
paper by Moretti et al.). We built a list of candidate of extended
sources, by checking for each candidate source if {\it detect} finds a
clusters of spurious sources on the diffuse emission, and/or if the
X-ray contours show extended emission. Then, we verified that a source
is actually extended, by comparing the source brightness profile with
the {\it XRT} PSF at the source position on the detector, using {\it
XIMAGE}. We find that the number of these clearly extended sources is $<
10\%$ and $< 9\%$ of the sample, at a detection significance level of
P$=$4$\times10^{-4}$ and P$=$2$\times10^{-5}$, respectively.  Finally
we refined the source position by the task {\it xrtcentroid} of
the XRTDAS package.

\subsection{Catalog reliability}

To evaluate the number of spurious sources corresponding to the chosen
probability threshold of 2$\times10^{-5}$, we simulated 45 XRT
fields, with the same characteristics (i.e. number of observations,
exposure times, R.A. and Dec. of the single pointings) of the fields,
which were randomly chosen among the 374 XRT fields.

The simulations were made up by an X-ray event simulator,
developed at the ASI Science Data Center (ASDC), already used for
missions like Beppo-SAX, Simbol-X, Nustar, Swift-XRT (see e.g. the
flow chart in Puccetti et al. 2009b, and a few examples of
applications in Puccetti et al. 2008, Fiore et al. 2008, 2009.). For
{\it Swift}-XRT we updated the ASDC simulator with the
calibration files distributed by
heasarc\footnote{http://heasarc.gsfc.nasa.gov/docs/heasarc/caldb/data/swift/xrt/}
(i.e. the vignetting function, the analytical function describing the
PSF and the response matrix files) and with the XRT background
described in Moretti et al. 2009. The simulated sources is randomly
drawn from the 0.5-2 keV X-ray number counts predicted by the AGN
population synthesis model by Gilli et al. (2007).

For each field we first simulated an observation with an exposure time
increased by a factor of 5 compared to the original value, to generate
a source list deeper than that of the original XRT field. This source
list was then used as input for each of the observations of the same
XRT field. Finally we summed all the observations of the same field,
as for the real fields (see Sec. 3) and applied the detector procedure
and the visual cleaning described in Sec. 4.1. We matched the input
and the detected source lists using a maximum likelihood algorithm
with maximum distance of 6 arcs, to find the most probable association
between an input source and an output detected source. By this
analysis, we find a total of 11 spurious sources in the 45 simulated
fields. Therefore we evaluated an average number of spurious sources
of 0.24 for each field in the three energy bands (S, H, and F) at the
probability threshold of 2$\times10^{-5}$.

\subsection{Count rates, fluxes}

\begin{figure}
\begin{center}
\includegraphics[angle=0,height=6truecm,width=6truecm]{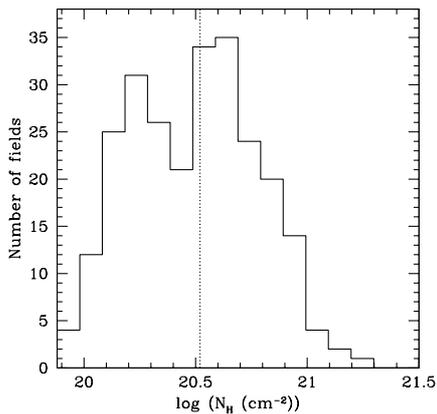}
\caption{Distribution of the Galactic hydrogen column density for the
254 HGL fields. The dotted line indicates the median value
3.3$\times$10$^{20}$cm$^{-2}$. }
\label{nhf}
\end{center}
\end{figure}

\begin{figure}
\begin{center}
\includegraphics[angle=0,height=6truecm,width=6truecm]{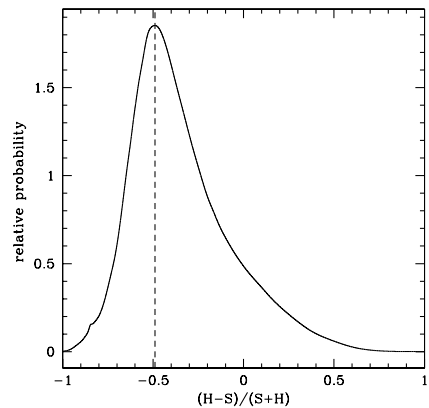}
\caption{Relative probability density distribution of the hardness ratio
(H-S)/(H+S) for the high Galactic-latitude sample. The dashed line indicates the most probable value of
hardness ratio, which corresponds to a power-law spectral model with
$\Gamma=1.8$, absorbed by an hydrogen column density
N$_H$$=$3.3$\times 10^{20}$ cm$^{-2}$.}
\label{gf}
\end{center}
\end{figure}

For a sample of 20 sources in a broad range of brightness (F flux in
the range 3.9$\times$10$^{-15}$$\div$1.3$\times$10$^{-14}$ erg
cm$^{-2}$ s$^{-1}$) and off-axis angles, we compared the count rates
evaluated using the {\it detect} algorithm with the count rates
measured from the spectra extracted using a radius of 20 arcs, which
corresponds to a f$_{psf}$$\sim$70\%-80\%, depending on energy and
off-axis angle. The count rates measured from the spectra were then
corrected for the f$_{psf}$ and the telescope vignetting, using the
appropriate response matrix. The average ratio between the count rates
given by the {\it detect} algorithm and those measured from the
spectra is 1.1$\pm$0.2, indicating a good consistency between the two
methods at 1$\sigma$ confidence level.  \\ For the high
Galactic-latitude sample ($\mid$b$\mid\ge$20, HGL catalog
hereinafter), in order to be consistent with other results present in
the literature, count rates estimated in the S, H and F band were
extrapolated to 0.5--2~keV, 2--10~keV and 0.5--10~keV fluxes,
respectively. To convert count rates into fluxes, we assumed that the
typical spectrum of the HGL sources is a simple power-law model
absorbed by the Galactic column density along the line of sight. We
chose to fix N$_H$ to the median of the Galactic N$_H$ of the HGL
fields, that is 3.3$\times 10^{20}$ cm$^{-2}$ with an interquartile
range of 1.4$\times 10^{20}$ cm$^{-2}$(see Fig. \ref{nhf}). We then fixed the spectral slope of the power-law model, to the most
probable value, according to the distribution of the hardness ratio,
defined as HR=(c$_H$-c$_S$)/($c_H$+c$_S$), where c$_S$ and c$_H$ are
the S and H count rates of the HGL sources detected in both the bands,
respectively. Following Mateos et al. (2008), we assume that each
source has an HR distributed as a Gaussian with mean value HR and
$\sigma$, the 68\% error on HR. We then calculated the integrated
probability by adding the probability density distributions of the HR
of each source (see Fig. \ref{gf}). We find that the most probable
value is HR$\simeq$-0.5, that for N$_H$=3.3$\times 10^{20}$ cm$^{-2}$,
corresponds to a photon spectral index $\Gamma$=1.8, assuming a power
law model\footnote{f$_E$$\propto$E$^{-\alpha}$ with
$\Gamma=\alpha+1$.}.

Count rates were converted to fluxes using the conversion factors
quoted in the first line of Table \ref{conversions}, which are
appropriate for a power law spectrum with photon spectral index
$\Gamma$=1.8, absorbed by a Galactic N$_H$$=$3.3$\times 10^{20}$
cm$^{-2}$. The major uncertainty in the estimate of the fluxes is due
to the variety of intrinsic spectra of the X-ray sources.
Moreover the average spectral properties are a function of
the observed flux (Brandt \& Hasinger 2005). To estimate this
uncertainty, we calculated the count rates to fluxes conversion
factors for power law spectra with $\Gamma=$1.4, and for absorbed
power law spectra with $\Gamma=$1.4 and 1.8, and N$_H=10^{22}$
cm$^{-2}$. The conversion factors are in ranges of
$\sim$1-1.3,$\sim$1.1-1.2 and $\sim$1.3-2.1, in the S, H and F band,
respectively (see Table~\ref{conversions}). The conversion factor for
the F band is more sensitive to the spectral shape than for the S and
H bands, because it is wider. 

For the low Galactic-latitude sources we used the same conversion
factors of the HGL sample, to convert count rates to fluxes.

\begin{table*}
\caption{Count rate to flux conversion factors for different spectral models.}
\begin{tabular}{lcccc}
\hline
$\Gamma$ & N$_H$       &  CF(F)$^a$ &   CF(S)$^b$  &  CF(H)$^c$ \\       
           & $10^{22}$ cm$^{-2}$ & cts s$^{-1}$ $/$10$^{-11}$ erg cm$^{-2}$ s$^{-1}$ &cts s$^{-1}$ $/$10$^{-11}$ erg cm$^{-2}$ s$^{-1}$ & cts s$^{-1}$ $/$10$^{-11}$ erg cm$^{-2}$ s$^{-1}$\\
\hline
1.8 & 0.033  & 3.641& 1.591& 8.090 \\
1.4 & 0.033  & 4.868 & 1.565 & 9.283\\
1.4 & 1      & 7.720 & 1.232 & 9.880\\
1.8 & 1      & 6.324 & 1.326 & 8.620 \\
\hline
\end{tabular}

$^a$energy conversion factor to convert the F band count rate into 0.5-10 keV flux assuming an absorbed power-law spectrum with hydrogen
column density N$_H$ and photon index $\Gamma$;

% using the formula  {\it Flux}$=$count rates$\times$CF$\times$10$^{-11}$ and appropriate f
$^b$same as $^a$, but to convert the S band count rate into the 0.5-2 keV flux;

$^c$same as $^a$, but to convert the H band count rate into the 2-10 keV flux.
\label{conversions}
\end{table*}

\subsection{Upper limits}

If a source is not detected in one band, we give a 90\%
upper limits to the source count rates and fluxes. The upper limits
are computed following Puccetti et al. 2009. If M is the number of
counts measured at the position of each source in a region of 16.5
arcs radius, which corresponds to a mean f$_{psf}$ of $\sim68\%$, B are the
background counts, evaluated by the background maps (see Sec. 3), and
$\sigma=$$\sqrt B$, the 90\% upper limit is defined as the number of
counts X that gives 10\% probability to observe M (or less) counts
equal to the Poisson probability:

\begin{equation}
P_{Poisson}=e^{-(X+B)}\sum_{i=0}^M { (X+B)^i \over i!}
\end{equation}

We solved Eq. 1 iteratively for a 10\% probability. The X upper
limits derived with Eq. 1 do not take into account the statistical
fluctuations on the expected number of background counts. In order to
take the background fluctuations into consideration, we used the
following procedure: if $\sigma(B)$ is the root mean square of B
(e.g., $\sigma(B)=\sqrt(B)$ for large B), we estimated the 90\% lower
limit on B as B(90\%)$=$B-1.282$\times\sigma$(B) \footnote{The
value 1.282 is the value appropriate for the 90\% probability (see
e.g., Bevington P.R. and K. Robinson 1992).} and, as a consequence,
the ``correct'' 90\% upper limit (Y) becomes $Y = X \times 1.282 \times \sigma$.

Vignetting corrected count rates limits for each source are obtained
by dividing the count upper limits by the net exposure time, reduced
by the vignetting at the position of each source, as in the
corresponding exposure maps (see Sec. 3) and by correcting for the  f$_{psf}$.

\subsection{Positional error}

The total positional uncertainty results from the combination of the
statistical uncertainty (i.e. $\sigma_{stat}$), that depends on the
instrumental PSF at the position of the source and is inversely
proportional to the source counts, and of the uncertainty on the XRT
aspect solution (i.e. $\sigma_{asp}$). The total
positional uncertainty is:

\begin{equation}
err_{pos} = \sqrt{ {\sigma_{stat}}^2 + {\sigma_{asp}}^2}
\end{equation}

We evaluated the positional errors at 68\% and 90\%.  The
$\sigma_{stat}$ at 68\% level confidence are evaluated by dividing the
PSF$_{radius}$ corresponding to a mean f$_{psf}$ of 68\% (i.e
$\sim$16.5 arcs) to the square root of the background subtracted
source counts from aperture photometry, following Puccetti et
al. (2009).\\ The aperture photometry values are derived from the
total event data for each field. To extract source counts, circular
regions centered on each source with a 16.5 arcs radius,
corresponding to a mean f$_{psf}$ of 68\% for different
off-axis angles and energies, are used. The background counts are
extracted from the background maps calculated as described in
Sec. 3.\\The $\sigma_{stat}$ at 90\% level confidence are evaluated
following the formula by Hill et al. (2004):
R$\times$counts$^{-0.48}$, with R$=$22.6 arcs and counts are the
background subtracted source counts corresponding to a mean
f$_{psf}$ of $\sim$80\%.

We cross-correlated the XRT catalog cut at a significance level of
P$=$2$\times$10$^{-5}$ and with source count rate equal or greater than
0.001 ct/s, with the SDSS optical galaxy catalog to find the mean
$\sigma_{asp}$ at 68\% and 90\% confidence level. For the
cross-correlation, we used a match radius of 10 arcs, and a source
positional uncertainty of $\sqrt{ {\sigma_{stat68\%}}^2 +
{\sigma_{asp68\%}}^2}$ and $\sqrt{ {\sigma_{stat90\%}}^2 +
{\sigma_{asp90\%}^2}}$, varying $\sigma_{asp68\%}$ and
$\sigma_{asp90\%}$ to obtain that the XRT sources with an optical
counterpart are 68\% and 90\%, respectively. In this way we find that
the mean $\sigma_{asp}$ at 68\% and 90\% are 2.05 arcs and 3.55
arcs, respectively. The values of $\sigma_{asp}$ are consistent
with previous results by Moretti et al. (2006).

\begin{figure*}
\begin{center}
\begin{tabular}{cc}

\includegraphics[width=7truecm]{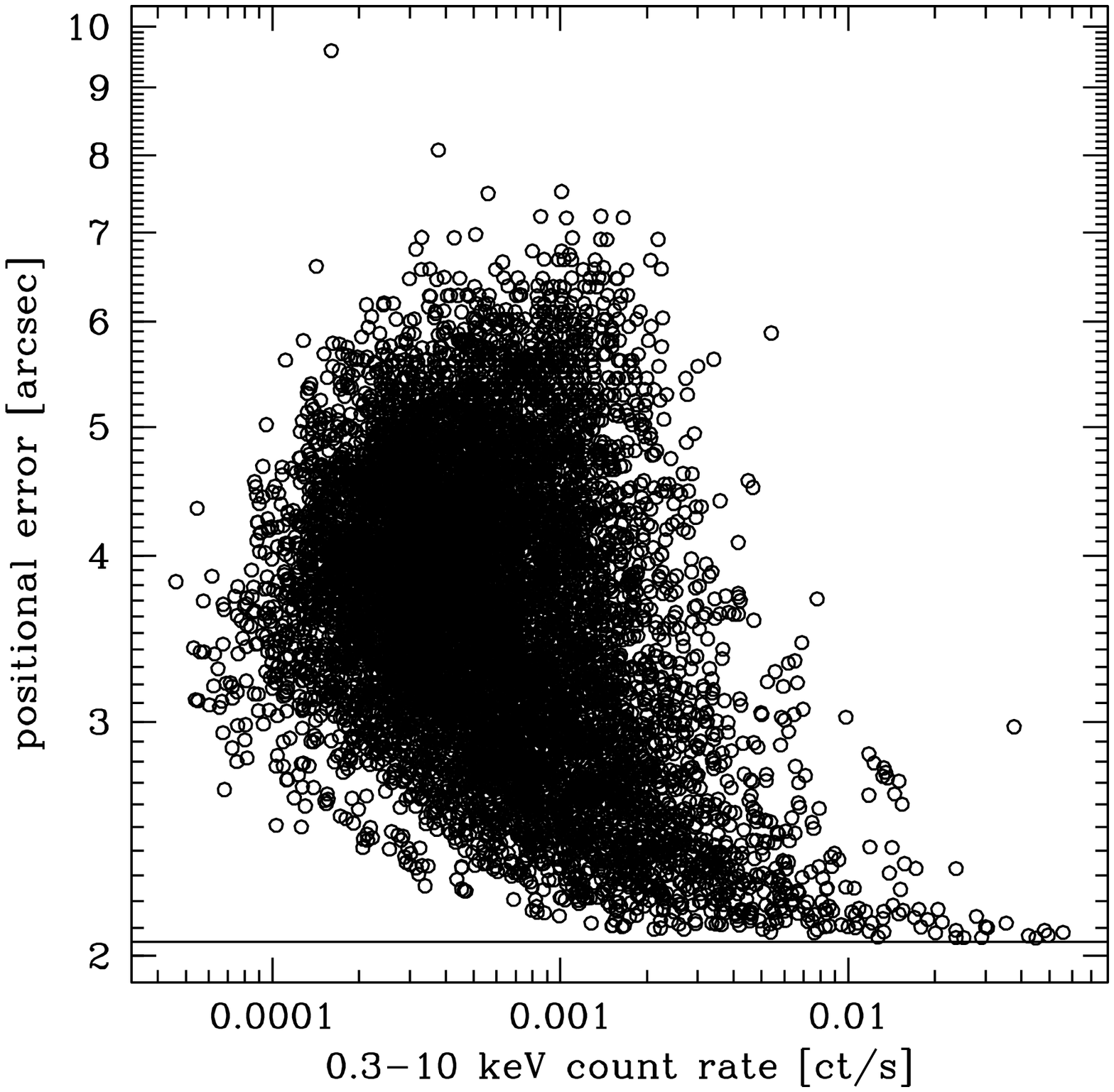}
\includegraphics[width=7truecm]{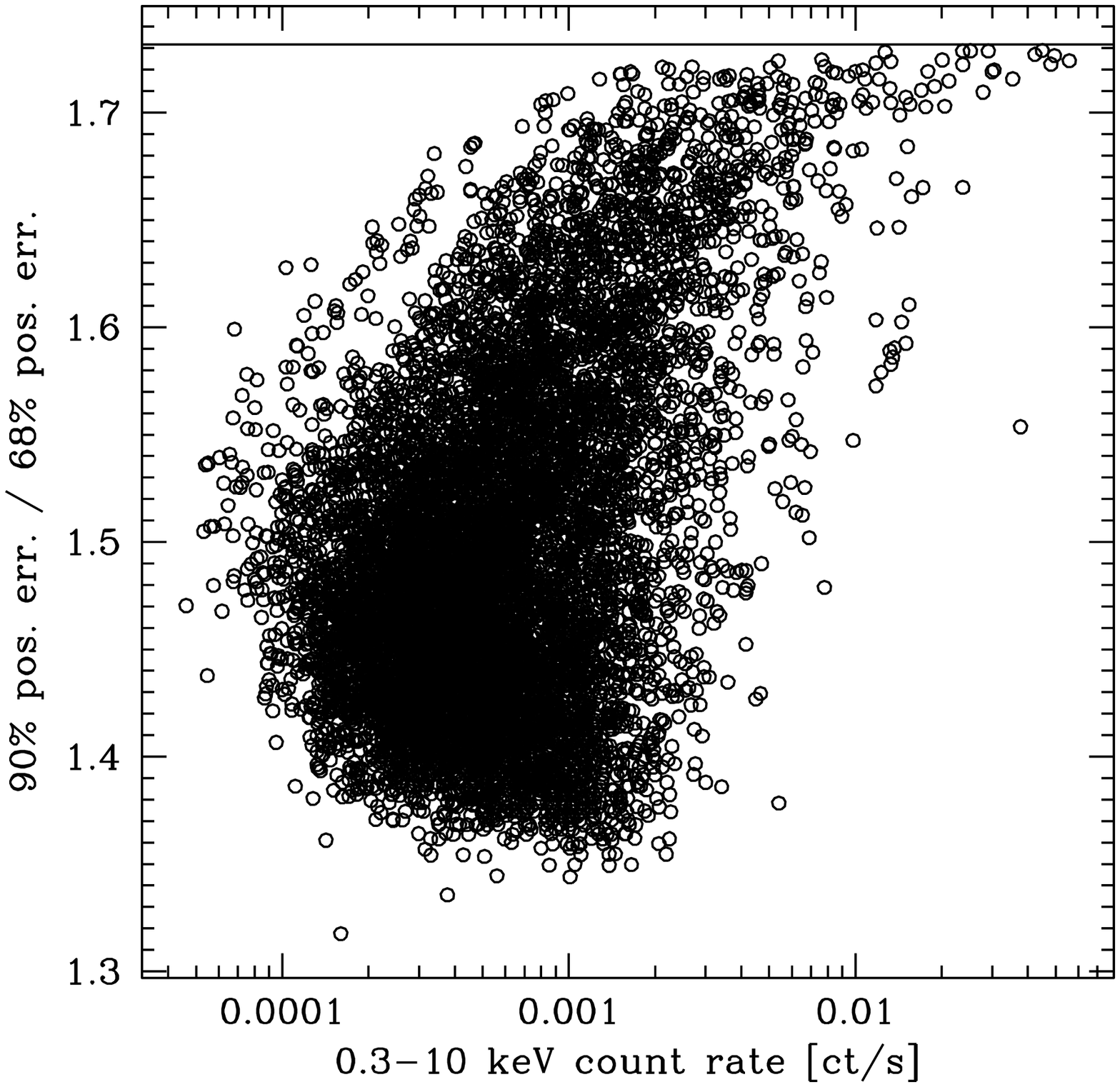}
\end{tabular}
\end{center}
\caption{{\it Left panel}: the 68\% positional errors vs. the F band
count rates, the solid line indicates the case with null statistical uncertainty ($\sigma_{stat}=0$). {\it
Right panel:} ratio between the 90\% positional error and the 68\%
positional error vs. the F band count rates, the solid line indicates
the case with null statistical uncertainty ($\sigma_{stat}=0$)}.
\label{plerr}
\end{figure*}

The left panel of Fig.~\ref{plerr} shows the 68\% positional errors as
a function of the F band count rates, the solid line indicates the
case in which the positional errors are exclusively due to
$\sigma_{asp}$. The right panel of Fig.~\ref{plerr} shows the ratio
between the 90\% positional error and the 68\% positional error
vs. the F band count rates, the solid line is the case in which
$\sigma_{stat}$ is equal to zero. We note that the positional error
ratio is not Gaussian (i.e. equal to $\sim$1.65), probably due to the
XRT PSF shape, which is not Gaussian.

\subsection{Source confusion}

In order to estimate the effects of source confusion on the HGL sample, 
we evaluated the probability P of finding two sources with flux F$_{\rm x}$ equal
or higher than a flux threshold (F$_{\rm x}$$_{min}$) at a distance closer
than the minimum angular separation $\theta_{min}$, following:

\begin{equation}
P(<\theta_{min})=1-e^{-\pi N \theta^{2}_{min}}
\end{equation}

where $\theta_{min}$ is set to twice the typical size of the source
cell detection (i.e. 4 original pixel), and N is the number counts at
F$_{\rm x}$$_{min}$, evaluated by the X-ray number counts of C-COSMOS (Elvis et
al. 2009).

The probability of finding two sources with flux higher than
F$_{\rm x}$$_{min}$= 2$\times10^{-15}$ erg cm$^{-2}$ s$^{-1}$ and F$_{\rm x}$$_{min}$=
1.1$\times10^{-14}$ erg cm$^{-2}$ s$^{-1}$, for the S, and H bands,
corresponding to a sky coverage of $\sim$2.2 square degrees (i.e. 10\%
of the HGL sky coverage), is only 4.6\% and 2.3\% for the S and H
bands, respectively. These probabilities increase to 9\% and 7.6\% for
fluxes corresponding to the faintest detected sources in the two
bands.

\subsection{CDFS: {\it Swift}-XRT vs Chandra}

\begin{figure}
\includegraphics[angle=0,height=7truecm,width=7truecm]{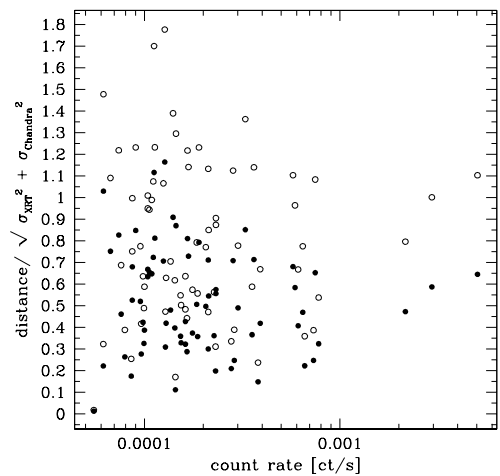}
\caption{Ratio between the distance of the nearest {\it Chandra}
source to each XRT source and the maximum radius
$\sqrt{{\sigma_{XRT}(68\%)}^2 +{\sigma_{\it Chandra}}^2}$ (open dots)
 as well as $\sqrt{{\sigma_{XRT}(90\%)}^2 +{\sigma_{\it Chandra}}^2}$ (solid
dots) as a function of the count rates in the F or S or H band (see
Sec. 4.7).  }
\label{xrtchpos}
\end{figure}

\begin{figure*}
\begin{center}
\begin{tabular}{cc}

\includegraphics[width=7truecm]{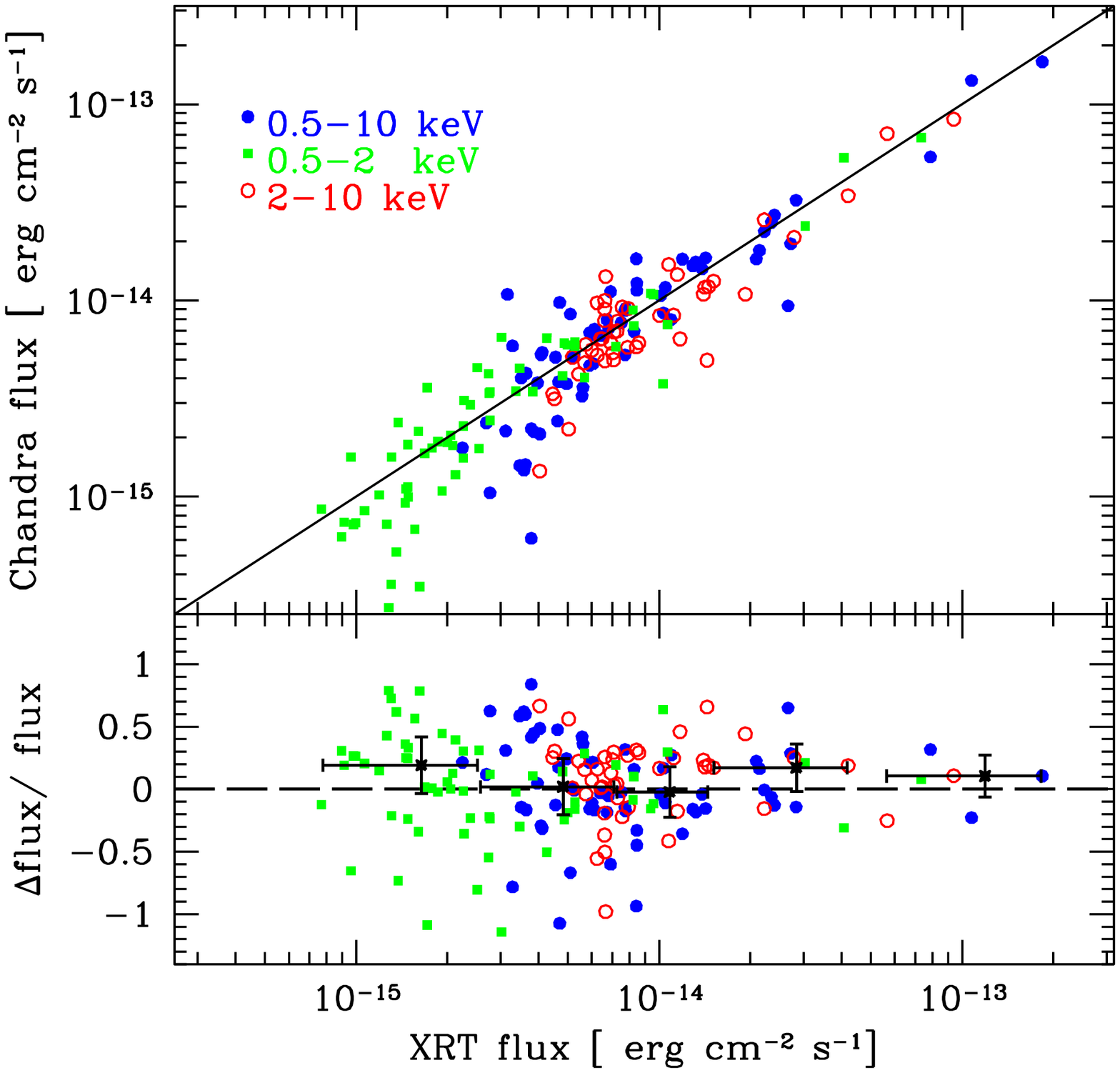}
\includegraphics[width=7truecm]{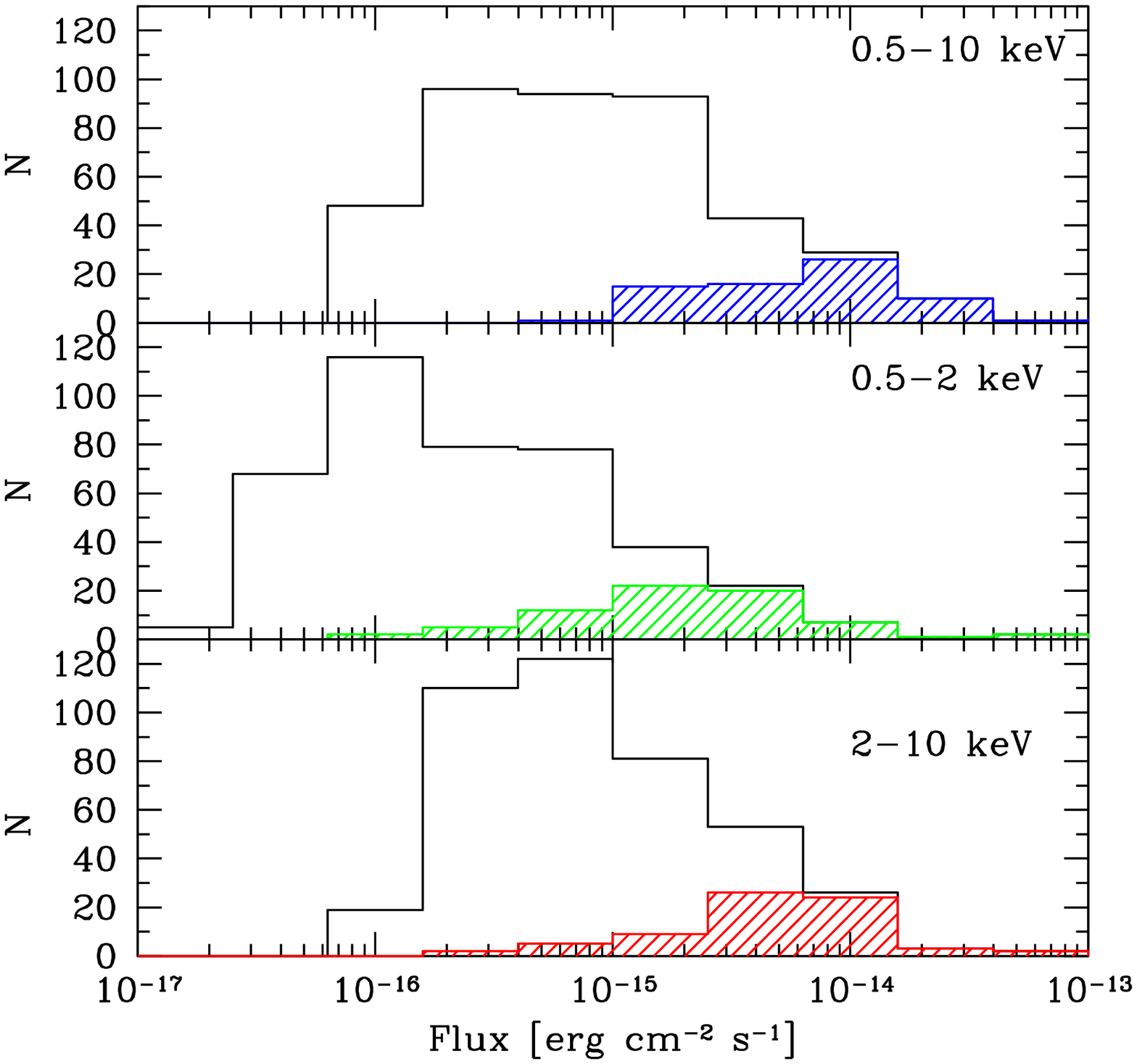}
\end{tabular}
\end{center}
\caption{{\it Left top panel}: the CDFS {\it Chandra} source fluxes
 vs. the XRT counterpart fluxes in the three energy band: 0.5-10 keV
 (blue solid dots), 2-10 keV (red open dots) and 0.5-2 keV (green solid
 squares). The solid line is the exact match between the {\it Chandra}
 and XRT fluxes. {\it Left bottom panel}: ratio between the
 relative difference of the XRT and {\it Chandra} fluxes vs. the XRT
 fluxes. The dashed line indicates the exact match between the {\it
 Chandra} and XRT fluxes, the stars are the mean ratios in each flux
 bin, with 1$\sigma$ uncertainties. {\it Right panel}: the empty
 histogram represents the flux distribution of the {\it Chandra}
 sources and the shaded histogram represents the flux distribution of
 the {\it Chandra} sources with an XRT counterpart, in the three
 energy band: 0.5-10 keV, 0.5-2 keV, 2-10 keV.}
\label{xrtchfx}
\end{figure*}

We applied the data cleaning, the source detection and source
characterization described above, to the CDFS XRT data and compared
the resulting CDFS XRT catalog, cut to a significance level of
2$\times 10^{-5}$ (see Sec. 4.2), to the {\it Chandra} catalog (Luo et
al. 2008). We found that 71 out of 72 XRT sources are within the {\it
Chandra} field. We matched the two CDFS catalogs using for each source
either the error circle given by the sum of the squares of the XRT
positional error (i.e.~$\sigma_{XRT}(68\%)$ and $\sigma_{XRT}(90\%)$
at 68\% and 90\% level confidence, respectively) and {\it Chandra}
85\% level confidence positional error (i.e. $\sigma_{Chandra}$) or a
fixed distance conservatively of 10 arcs.  Fig. \ref{xrtchpos} shows the ratio
between the distance of the nearest {\it Chandra} source to each XRT
source and the maximum radius $\sqrt{{\sigma_{XRT}(68\%)}^2
+{\sigma_{\it Chandra}}^2}$ as well as the maximum radius $\sqrt{{\sigma_{XRT}(90\%)}^2
+{\sigma_{\it Chandra}}^2}$ as a function of the count rates in the F
band, if the source is detected in the F band, otherwise in the S
band, otherwise in the H band. We find that the $\sim$80.2\% and the
$\sim$95.8\% of the XRT CDFS sources have a {\it Chandra} counterpart,
using the 68\% and 90\% level confidence XRT positional errors,
respectively. Three XRT sources have a marginal {\it Chandra}
detection at distance less than 6.5 arcs. Five XRT sources have two
{\it Chandra} counterparts inside the error circle, which corresponds
to $\sim$7\% source confusion at a flux limit of $\sim$1.2$\times
10^{-15}$ and $\sim$4$\times 10^{-15}$ erg cm$^{-2}$ s$^{-1}$ in the S
and H band, respectively. This percentage of source confusion is fully
consistent with the estimate in Sec. 4.6. We then compared the XRT and
{\it Chandra} fluxes in all the three bands 0.5-10 keV, 2-10 keV and
0.5-2 keV. We find good flux consistency (see left panel of
Fig. \ref{xrtchfx}), regardless of source variability. Actually the
faintest XRT fluxes, near the flux limit, and the XRT fluxes around
3$\times 10^{-14}$ erg cm$^{-2}$ s$^{-1}$, although consistent at
1$\sigma$ confidence level with the {\it Chandra} fluxes, appear
systematically greater than the {\it Chandra} fluxes (see left bottom
panel of Fig. \ref{xrtchfx}). This trend for the faintest XRT sources
is probably due to the Eddington bias, while for the brightest sources the
statistics are too poor to permit a firm comparison. Finally the
right panel of Fig. \ref{xrtchfx} shows the comparison between the
flux distribution of the total {\it Chandra} catalog and the {\it
Chandra} source with XRT counterparts.

\section{The point-source catalog}

\begin{figure*}
\begin{center}
\begin{tabular}{cc}

\includegraphics[width=7truecm]{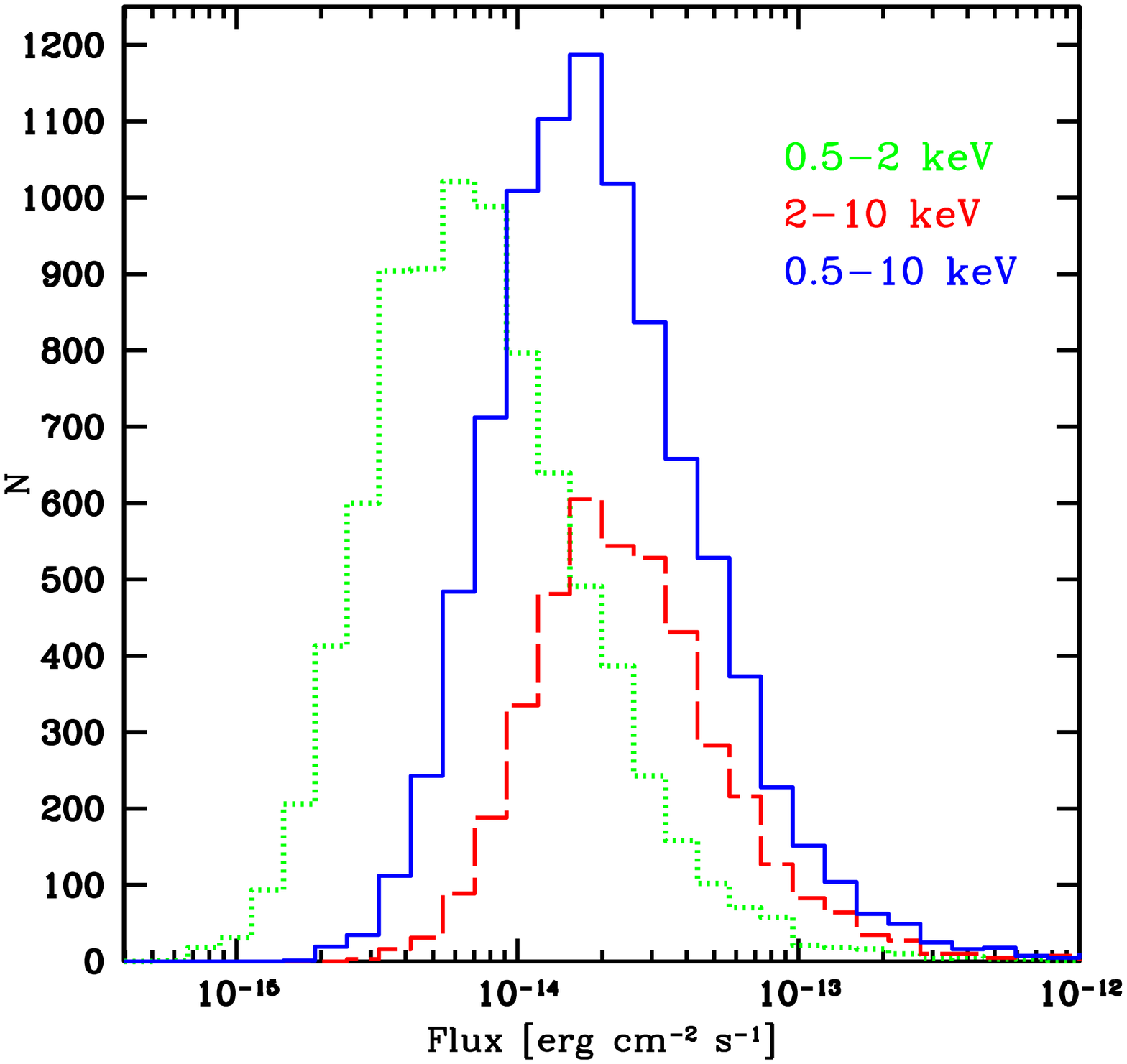}
\includegraphics[width=7truecm]{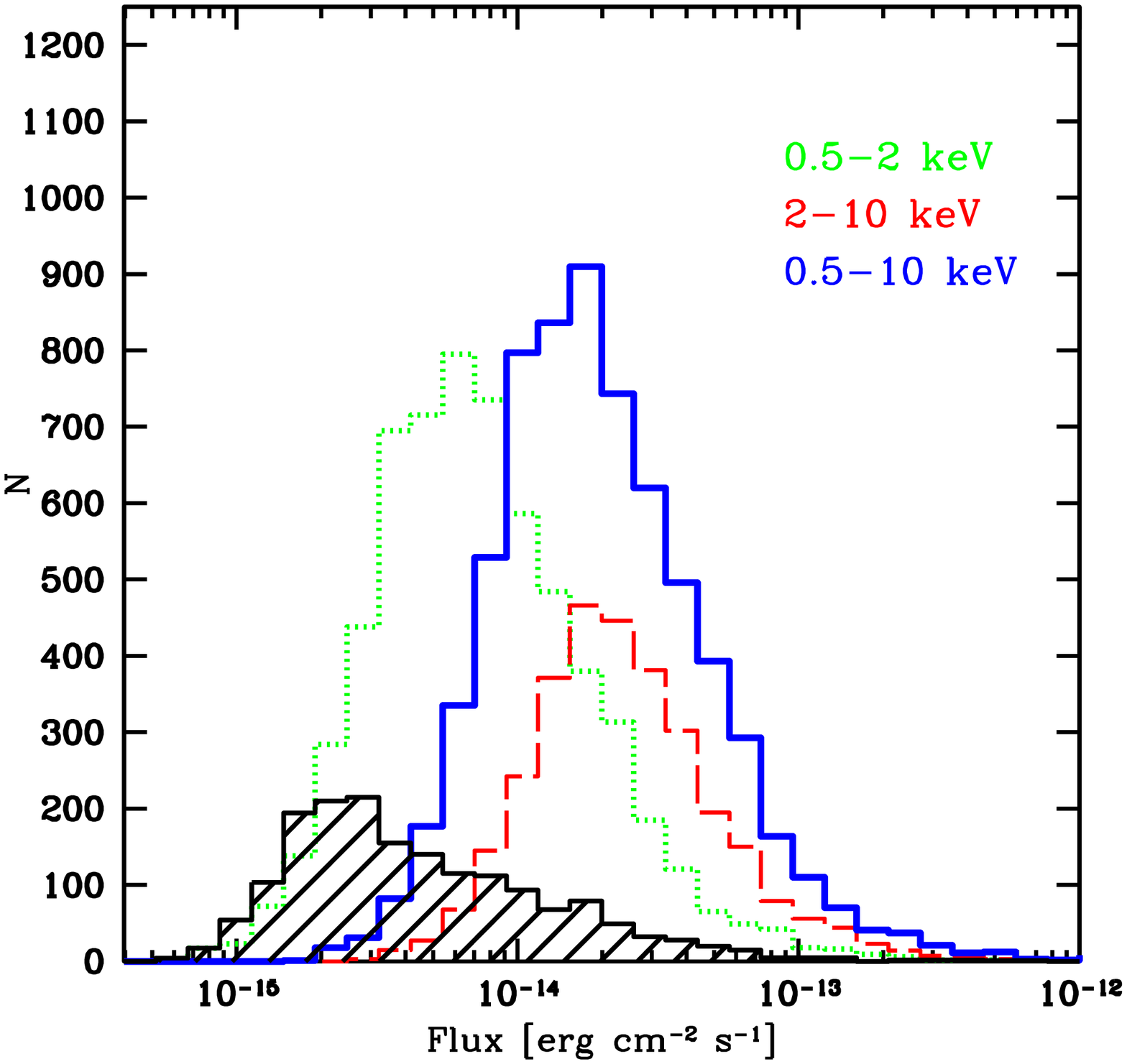}
\end{tabular}
\end{center}
\caption{{\it Left panel}: the flux distributions of those sources detected
in the S (green dotted histogram), H (red dashed histogram) and F
(blue solid histogram) band in the total sample. {\it Right panel}:
the flux distribution of those sources detected in the S (green dotted
histogram), H (red dashed histogram) and F (blue solid histogram) band
in the HGL sub-sample. The black shaded histogram represents the flux
distribution of the 0.5-7 keV C-COSMOS sources (Elvis et al. 2009).}
\label{fxhisto}
\end{figure*}

The {\it detect} tool was run on the three bands S, H and F. Table
\ref{sourcesattwodetml} gives the numbers of sources detected in each
band with different significance level
in both total and HGL catalog. We produced a unique catalog merging
the individual S, H and F lists, using a matching radius of 6
arcs. We retained reliable sources, i.e. those with a significance
level of being spurious $\le$2$\times$10$^{-5}$ in at least one band,
to limit the number of spurious detections to $\sim$0.24 for field.
The final total and HGL catalog contain
9387 and 7071 sources, respectively. Table \ref{sourcesinbands}
reports the numbers of total and HGL catalog sources detected in three
bands, two bands, or in only one band. Fig. \ref{fxhisto} shows the
flux distributions of the total sample (left panel) and of the HGL
sample (right panel). We detect sources in the 0.5-2 keV and 2-10 keV
bands down to flux limits of $\sim 7\times 10^{-16}$ \cgs and $\sim
4\times 10^{-15}$ \cgs, respectively. In comparison with a typical deep
contiguous medium area survey, like C-COSMOS (Elvis et al. 2009, see
Fig.\ref{fxhisto}) the advantage of the SwiftFT is the definitely larger
number of sources and the wider flux coverage, despite a slightly
higher flux limit.

\begin{table}[h]
\footnotesize
\caption{Number of sources detected in each band at the two adopted probability thresholds.}
\begin{center}
\begin{tabular}{lcccc }
\hline
Band & N$^a$ & N$_1$$^b$ & N$_{HGL}$$^c$ & N$_{HGL}$$_1$$^d$ \\
\hline
F&  8719 & 880 & 6596 & 639\\
S&  7925 & 684 & 6062 & 501 \\
H&  3791 & 436 & 2819 & 337 \\
\hline
\end{tabular}
\end{center}
$^{a}$ number of detected sources with detection  significance level $\leq$ 2$\times10^{-5}$;

$^{b}$number of detected sources with detection  significance level: 2$\times 10^{-5} \le $prob$\leq 4\times10^{-4}$;

$^{c}$ number of detected sources in the HGL fields, with detection significance level $\leq$2$\times10^{-5}$;

$^{d}$ number of detected sources in the HGL fields, with detection significance level: 2$\times10^{-5}\le$prob$\leq4\times10^{-4}$.
\label{sourcesattwodetml}
\end{table} 

\begin{table}[h]
\begin{center}
\caption{Number of detected sources in the total SwiftFT catalog, with sources having a detection significance level
$\le$2$\times$10$^{-5}$ in at least one band.}
\begin{tabular}{lcc}
\hline
Band & N$^{a}$& N$_{HGL}$$^{b}$\\
\hline
F & 8986  &   6797 \\
S & 8202  & 6253 \\
H &4120  & 3088\\
\hline 
F+S+H &3498 &2671\\
F+S & 4404 &3371\\
F+H &521  &354\\
F only & 563 &401\\
S only & 300 &211\\
H only &101 &63\\
\hline
\end{tabular}
\end{center}
%\end{small}
$^{a}$ number of detected sources in the total SwiftFT catalog;

$^{b}$ number of detected sources in the HGL catalog.
\label{sourcesinbands} 
 \end{table}

\subsection{Catalog description}

The full catalog is available on-line at this address
http://www.asdc.asi.it/xrtgrbdeep\_cat/.  Table \ref{cata} gives the
parameter descriptions of each source and Table \ref{cataex} gives ten
entries as an example.

\begin{table}
\caption{Source parameters in the catalog.}
\begin{center}
\begin{tabular}{|rrl|}
\hline
Column & Parameter & Description\\
\hline
1 & NAME & source name: prefix SWIFTFTJ, following the standard IAU convention\\
2 & RA & {\it Swift}-XRT Right Ascension in hms in the J2000  coordinate system.\\
3 & DEC & {\it Swift}-XRT Declination in hms in the J2000  coordinate system.\\
4 & pos\_err & Positional error at 68\% confidence level in arcs\\
5 & pos\_err & Positional error at 90\% confidence level in arcs\\
6 & X & X pixel coordinate \\
7 & Y & Y pixel coordinate \\
8 & Target name & XRT field \\
9 & START-DATE & Start time of the field observations in year-month-day h:m:s\\
10 & END-DATE & End time of the field observations in year-month-day h:m:s\\
11 & ON-TIME & Total on-time in sec\\
\hline
12 & f\_rate       & 0.3--10~keV count rate or 90\% upper limit in counts/sec\\
13& f\_rate\_err  & 1$\sigma$ 0.3--10~keV count rate error in counts/sec, in case of upper limits is set to -99\\
14 & f\_flux       & 0.5--10~keV Flux or 90\% in erg~cm$^{-2}$s$^{-1}$  \\
15 & f\_flux\_err  & 1$\sigma$ 0.5--10~keV Flux error in erg~cm$^{-2}$s$^{-1}$, in case of upper limits is set to -99\\
16 & f\_prob  & 0.3--10~keV detection probability\\
17 & f\_snr       & 0.3--10~keV S/N \\
18 & f\_exptime   & 1.5~keV exposure time in ks from the exposure maps \\
\hline
19& s\_rate       & 0.3--3~keV count rate or 90\% upper limit in counts/sec\\
20& s\_rate\_err  & 1$\sigma$ 0.3--3~keV count rate error counts/sec, in case of upper limits is set to -99\\\
21& s\_flux       & 0.5--2~keV Flux or 90\% upper limit in erg~cm$^{-2}$s$^{-1}$\\
22& s\_flux\_err  & 1$\sigma$ 0.5--2~keV Flux error in erg~cm$^{-2}$s$^{-1}$, in case of upper limits is set to -99\ \\
23 & s\_prob  & 0.3--13~keV detection probability\\
24 & s\_snr       & 0.3--3~keV S/N \\
25 & s\_exptime   & 1~keV exposure time in ks from the exposure maps  \\
\hline
26& h\_rate       & 2--10~keV count rate or 90\% upper limit in counts/sec\\
27& h\_rate\_err  & 1$\sigma$ 2--10~keV count rate in counts/sec, in case of upper limits is set to -99\\
28& h\_flux       & 2--10~keV Flux or 90\% upper limit in erg~cm$^{-2}$s$^{-1}$\\
29& h\_flux\_err  & 1$\sigma$ 2--10~keV Flux error in erg~cm$^{-2}$s$^{-1}$, in case of upper limits is set to -99\\
31 & h\_snr  & 2--10~keV detection probability\\
30 & h\_prob       & 2--10~keV S/N \\
32& h\_exptime    & 4.5~keV exposure time in ks from the exposure maps  \\
\hline
33 & hr  & hardness ratio = (h\_rate-s\_rate)/ (h\_rate+ s\_rate)/ \\
34 & ehr & 1$\sigma$  hardness ratio error evaluated with the error propagation formula (see e.g. Bevington 1992)\\
35 & off-axis & distance from the field median center in arcmin \\
36 & N$_H$ & Galactic hydrogen column density in cm$^{-2}$\\
\hline
\end{tabular}
\end{center}
\label{cata}
\end{table}

\begin{landscape}
\begin{table}
\caption{Catalog template}
\label{cataex}
\begin{tabular}{lccccccccccccccccccccccccccccccccccccccccccccccccccccccc}
\hline \hline 
NAME & RA & DEC & pos\_err68 & pos\_err90 & X & Y & Target name & START-DATE  & END-DATE  & ON-TIME  \\
\hline\hline 
SWIFTFTJ000234-5301.1 & 00 02 34.6 & -53 01 10.2 & 2.31 & 3.9 & 747.4 &  430.4& GRB070110  & 2007-01-10 07:27:08 & 2007-02-05 23:59:58 & 330057 \\
SWIFTFTJ000238-5255.9 & 00 02 38.0 & -52 55 54.1 & 2.55 & 4.1 & 734.5 &  564.5& GRB070110  & 2007-01-10 07:27:08 & 2007-02-05 23:59:58 & 330057 \\
SWIFTFTJ000239-5301.6 & 00 02 39.1 & -53 01 39.6 & 2.68 & 4.3 & 730.2 &  417.9& GRB070110  & 2007-01-10 07:27:08 & 2007-02-05 23:59:58 & 330057 \\
SWIFTFTJ000243-5259.3 & 00 02 43.3 & -52 59 22.9 & 3.93 & 5.8 & 713.9 &    476& GRB070110  & 2007-01-10 07:27:08 & 2007-02-05 23:59:58 & 330057 \\
SWIFTFTJ000252-5259.5 & 00 02 52.8 & -52 59 30.6 & 3.47 & 5.2 & 677.9 &  472.8& GRB070110  & 2007-01-10 07:27:08 & 2007-02-05 23:59:58 & 330057 \\
SWIFTFTJ000254-5250.9 & 00 02 54.7 & -52 50 54.4 & 4.28 & 6.2 & 670.9 &  691.8& GRB070110  & 2007-01-10 07:27:08 & 2007-02-05 23:59:58 & 330057 \\
SWIFTFTJ000255-5253.8 & 00 02 55.3 & -52 53 51.7 & 3.05 & 4.7 & 668.4 &  616.6& GRB070110  & 2007-01-10 07:27:08 & 2007-02-05 23:59:58 & 330057 \\
SWIFTFTJ000258-5301.3 & 00 02 58.0 & -53 01 19.1 & 3.56 & 5.3 & 657.9 &  426.8& GRB070110  & 2007-01-10 07:27:08 & 2007-02-05 23:59:58 & 330057 \\
SWIFTFTJ000300-5259.9 & 00 03 00.9 & -52 59 54.8 & 3.59 & 5.3 &  647  &  462.6& GRB070110  & 2007-01-10 07:27:08 & 2007-02-05 23:59:58 & 330057 \\
SWIFTFTJ000302-5301.0 & 00 03 02.9 & -53 01 03.6 & 3.82 & 5.6 &  639  &  433.5& GRB070110  & 2007-01-10 07:27:08 & 2007-02-05 23:59:58 & 330057 \\
\hline\hline 
\end{tabular}
\begin{tabular}{lccccccccccccccccccccccccccccccccccccccccccccccccccccccc}
f\_rate & f\_rate\_err & f\_flux & f\_flux\_err & f\_prob & f\_snr  & f\_exptime & s\_rate   & s\_rate\_err & s\_flux  & s\_flux\_err & s\_prob & s\_snr & s\_exptime \\
\hline \hline
 0.00124 &     9.5e-05  &  4.515e-14 &   3.459e-15 &           0  &      13.02&    2.705e+05&      0.00108   &   7.8e-05  &  1.718e-14 &   1.241e-15 &           0  &      13.84 &   2.717e+05   \\   	
   0.00063 &     6.4e-05  &  2.294e-14 &    2.33e-15 &           0  &      9.899&    2.819e+05&      0.0006    &     6e-05  &  9.546e-15 &   9.546e-16 &           0  &      9.963 &   2.829e+05  \\    						
  0.000432 &     5.5e-05  &  1.573e-14 &   2.003e-15 &           0  &      7.827&    2.772e+05&      0.00038   &   5.3e-05  &  6.046e-15 &   8.432e-16 &           0  &      7.115 &   2.783e+05  \\  
  0.000136 &     3.9e-05  &  4.952e-15 &    1.42e-15 &   2.553e-06  &      3.504&    2.884e+05&    0.0002099   &       -99  &   3.338e-15 &         -99 &           0  &          0 &   2.892e+05   \\   					
    0.000175 &     4.2e-05  &  6.372e-15 &   1.529e-15 &   4.673e-09  &      4.213&    3.018e+05&     0.000139   &   3.6e-05  &  2.211e-15 &   5.728e-16 &   1.294e-08  &      3.915 &   3.024e+05  \\  						
   0.000105 &     3.6e-05  &  3.823e-15 &   1.311e-15 &   0.0001077  &      2.889&    2.535e+05&     9.56e-05   &   3.2e-05  &  1.521e-15 &   5.091e-16 &   3.073e-05  &      2.949 &   2.547e+05  \\   						
   0.000259 &     4.5e-05  &   9.43e-15 &   1.638e-15 &           0  &      5.821&    2.968e+05&     0.000179   &   3.8e-05  &  2.848e-15 &   6.046e-16 &   5.789e-13  &      4.646 &   2.976e+05 \\     
   0.000139 &     3.8e-05  &  5.061e-15 &   1.384e-15 &   4.414e-07  &       3.69&    3.048e+05&     8.83e-05   &     3e-05  &  1.405e-15 &   4.773e-16 &    3.91e-05  &      2.966 &   3.053e+05 \\      					
  0.000162 &       4e-05  &  5.898e-15 &   1.456e-15 &    2.04e-08  &      4.057&    3.128e+05&     0.000149   &   3.6e-05  &  2.371e-15 &   5.728e-16 &   8.602e-10  &      4.154 &   3.132e+05  \\      						
   0.000129 &     3.7e-05  &  4.697e-15 &   1.347e-15 &   1.832e-06  &      3.526&    3.112e+05&    0.0005341   &       -99  &  8.499e-15 &         -99 &           0  &          0 &   3.116e+05  \\                                    
\hline \hline
\end{tabular}
\begin{tabular}{lccccccccccccccccccccccccccccccccccccccccccccccccccccccc} 
h\_rate & h\_rate\_err & h\_flux  & h\_flux\_err  & h\_prob & h\_snr  & h\_exptime  &  hr & ehr &  off$-$axis & N$_H$\\
\hline\hline 
    0.000245 &     4.3e-05 &   1.982e-14 &   3.479e-15 &           0  &      5.676 &   2.534e+05& -0.6302 & 0.06722 &  9.075  &   1.59e+20\\	  
   0.0006697 &         -99 &   5.419e-14 &         -99 &           0  &          0 &   2.671e+05& -99 &  -99 &  8.48& 1.58e+20\\			  
    0.000131 &     3.4e-05 &    1.06e-14 &   2.751e-15 &   3.558e-08  &       3.86 &   2.617e+05& -0.4873 &  0.1232 &  8.61 &     1.59e+20 \\	  
    0.000126 &     3.3e-05 &   1.019e-14 &    2.67e-15 &   2.517e-08  &      3.887 &   2.772e+05& -99 &  -99 &  7.382  &     1.59e+20\\		  
  0.0004435 &         -99 &    3.588e-14 &         -99 &           0  &          0 &    2.943e+05& -99 &  -99 &  6.006  &     1.59e+20\\	       	   
   0.0002909 &         -99 &     2.353e-14 &         -99 &           0  &          0 &   2.37e+05&  -99 &  -99 &  9.328  &     1.57e+20\\	       	  
    0.000144 &     3.3e-05 &   1.165e-14 &    2.67e-15 &    5.49e-11  &      4.376 &   2.863e+05& -0.1084 &  0.1558 &  7.115  &    1.58e+20\\	  
   0.0004109 &         -99 &   3.325e-14 &         -99 &           0  &          0 &   2.972e+05& -99 &  -99 &  5.908  &     1.6e+20\\		  
 0.0001906 &         -99 &   1.541e-14 &         -99 &           0  &           0 &    3.076e+05& -99 &  -99 &  4.933  &     1.6e+20\\	       	   
     7.2e-05 &     2.6e-05 &   5.825e-15 &   2.103e-15 &   0.0001587  &      2.802 &   3.054e+05& -99 &  -99 &  5.134  &   1.6e+20 \\              
\hline \hline
\end{tabular}

The columns are described in Table \ref{cata}.

\end{table}
\end{landscape}

\section{The high Galactic-latitude  ($\mid$b$\mid\ge$20 deg) catalog. }

\subsection{Survey sensitivity}

Telescope vignetting and changes in the PSF size (i.e. the background
counts) induce a sensitivity decrease toward the outer regions of the
detector. This effect, however, is not prominent in XRT, thanks to its
PSF and vignetting, that are approximately constant with the distance
from the center of the field of view. To evaluate survey
sensitivity in the F, S and H band, we followed the analytical
method, used for the case of ELAS-S1 mosaic (Puccetti et al. 2006 and
references therein). In this procedure, for each field in each
original pixel, we evaluated the minimum number of counts L, needed to
exceed the fluctuations of the background, assuming Poisson statistics
with a threshold probability equal to that assumed to cut the catalog
(i.e.$=2\times 10^{-5}$, see Sect. 4.2), according to the following
formula:

\begin{equation}
P_{Poisson}=e^{-B}\sum_{k=L}^\infty{B^k \over k!}=2\times 10^{-5}
\end{equation}

where B is the background counts computed from the background maps in
a circular region centered at the position of each pixel and of radius
R. R corresponds to a mean f$_{psf}$ $\sim$26\%, which corresponds
to a radius of $\sim$2 pixels, consistent with the sliding cell size
used by {\it detect}. We solved Eq. 4 iteratively to calculate
L. The count rate limit, CR, at each pixel of each field is then
computed by:

\begin{equation}
CR= {L-B \over{ f_{psf} \times T}}
\end{equation}

where T is the total, vignetting-corrected, exposure time at
each pixel read from exposure maps. This procedure, is applied for the
S and H bands to produce sensitivity maps. CR are thus converted to
minimum detectable fluxes (limiting flux) using the defined count
rate--flux conversion factors for the S and H bands, respectively (see
Sect. 4.3).

\subsection{Sky coverage}

\begin{figure}
\includegraphics[angle=0,height=8truecm,width=8truecm]{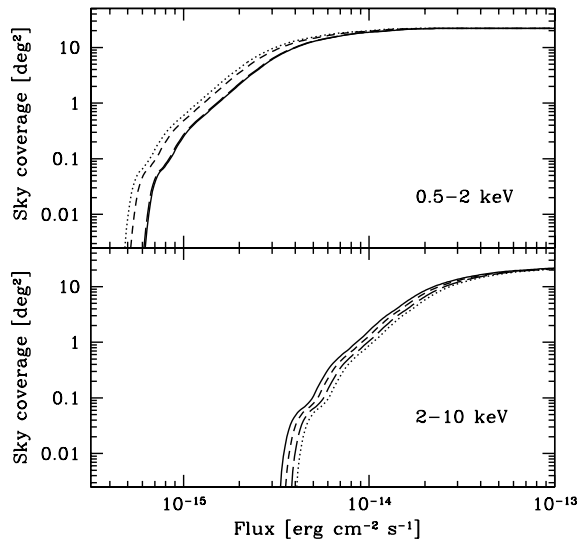}
\caption{The sky coverage calculated as in Sec. 6.2 for the 0.5-2 keV
({\it top panel}) and 2-10 keV ({\it bottom panel}) band. The {\em
solid} lines represent the sky coverages evaluated with the baseline
model (i.e. power-law spectra with $\Gamma=1.8$ absorbed by Galactic
N$_H=3.3 \times 10^{20}$ cm$^{-2}$). The {\em dotted} lines represent
the sky coverages for power-law spectra with $\Gamma=1.4$ absorbed by
Galactic N$_H=10^{22}$ cm$^{-2}$. The {\em short-dashed} lines
represent the sky coverages for power-law spectra with $\Gamma=1.8$
absorbed by N$_H= 10^{22}$ cm$^{-2}$. The {\em long-dashed} lines
represent the sky coverages for power-law spectra with $\Gamma=1.4$
absorbed by N$_H=3.3 \times 10^{20}$ cm$^{-2}$.  }
\label{skycov}
\end{figure}

``Sky coverage'' defines the area of the sky covered by a survey to a
given flux limit, as a function of the flux.  The sky coverage at a
given flux is obtained from the survey sensitivity, by adding up the
contribution of all detector regions with a given flux limit. Note
that we excluded a circular areas of radius 20 arcs
centered on the detected GRB. Figure \ref{skycov} plots the resulting
sky coverage in the S and H band.

The main sky coverage uncertainty is due to the unknown spectrum of
the sources near the detection limit. To estimate, at least roughly,
this uncertainty, we calculated the sky coverage also for power law
spectra with $\Gamma=1.4$, and for absorbed power law spectra with
$\Gamma=1.4,1.8$ and N$_H=10^{22}$ cm$^{-2}$, in addition to the
baseline case (see Fig. \ref{skycov}).

\subsection{The X-ray number counts}

The integral X-ray number counts are evaluated using the following
equation:

\begin{equation}
N(>{\it S})=\sum_{i=1}^{N_{\it S}}{ 1 \over {\Omega_i}} deg^{-2}
\end{equation}

where N$_{\it S}$ is the total number of detected sources with fluxes
higher than {\it S}, and $\Omega_i$ is the sky coverage at the flux of the
i-th source, evaluated as described in Sec. 6.2.

\begin{figure*}[h]
\begin{tabular}{cc}
\includegraphics[angle=0,height=8truecm,width=8truecm]{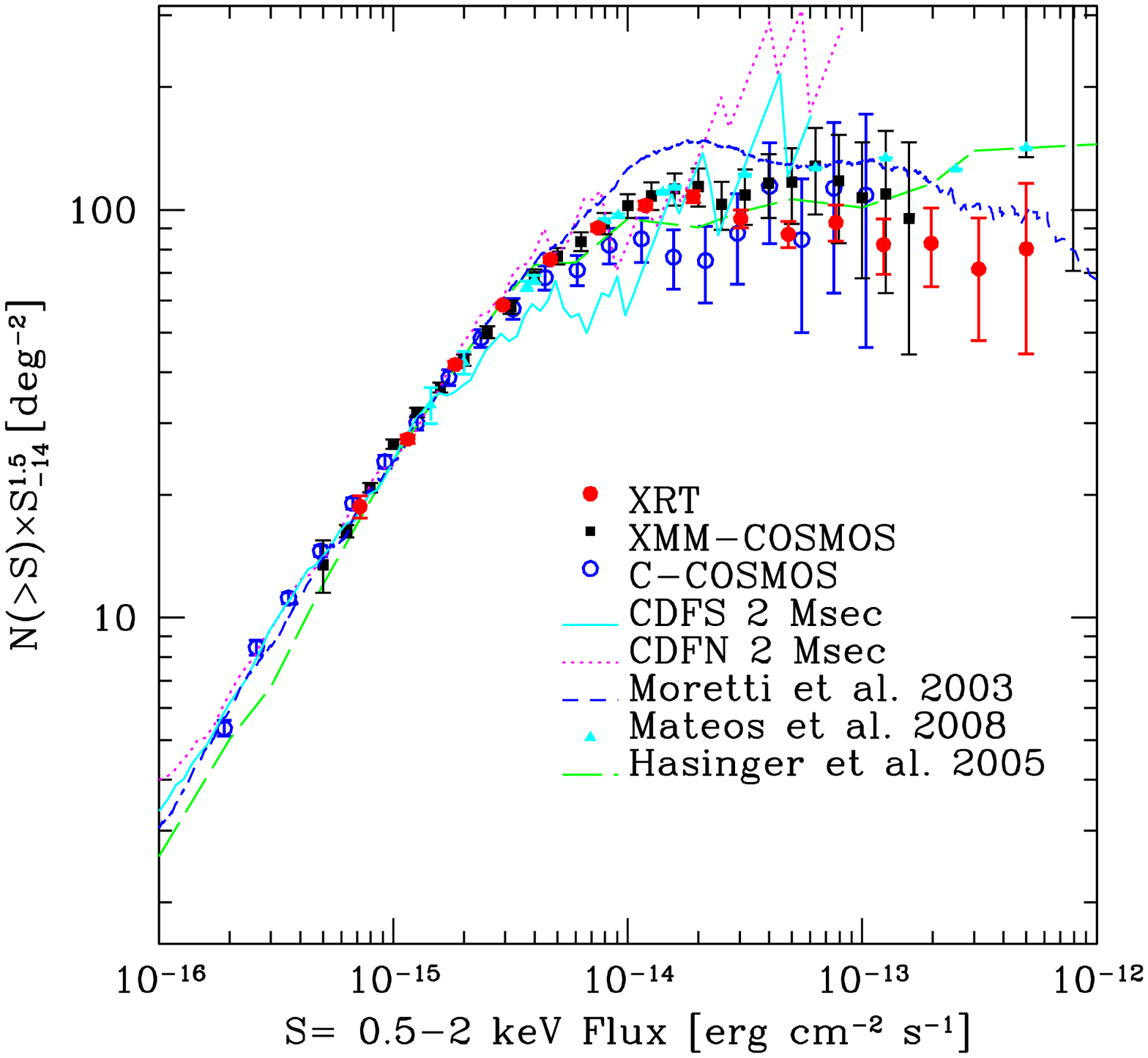}
\includegraphics[angle=0,height=8truecm,width=8truecm]{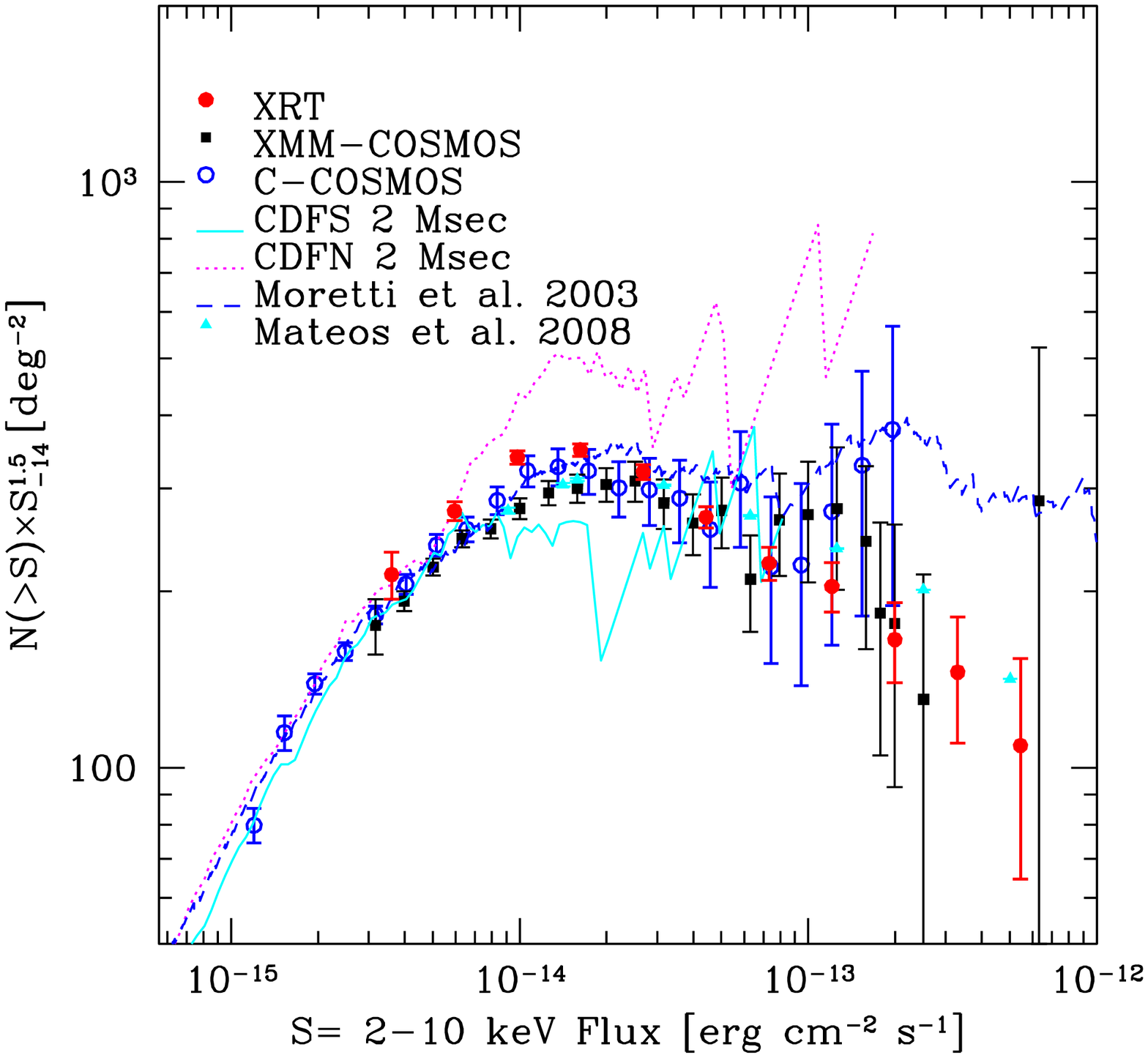}
\end{tabular}
\caption{Cumulative number counts normalized to the Euclidean slope
  (multiplied by {\it S}$^{1.5}_{-14}$) for HGL sample with detection
  significance level $\le$2$\times$10$^{-5}$ in the soft band
  (0.5--2~keV, red filled circles, left panel) and in the hard band
  (2--10~keV, red filled circles, right panel). Other symbols
  represent: the C-COSMOS curve (blue open circles, Elvis et
  al. 2009), the XMM-COSMOS curve (black filled squares, Cappelluti et
  al. 2009), the Moretti et al. (2003) compilation (blue dashed line),
  the soft band curve of Hasinger et al. (2005; green dashed line),
  the CDF-N (magenta dotted line, Alexander et al. 2003) and the CDF-S
  (cyan solid line, Luo et al. 2008) curves, the Mateos et al. (2008)
  compilation (cyan solid triangles). }
\label{lnls}

\end{figure*}

The cumulative number counts in the 0.5-2 keV and 2-10 keV bands are
reported in Table 7, while Fig. \ref{lnls} shows the cumulative number
counts normalized to the Euclidean slope ( multiplied by {\it
S}$^{1.5}$); Euclidean number counts would correspond to horizontal
lines in this representation.  Comparing the XRT number counts in the
largest possible flux range, we show in Fig. \ref{lnls} results from
other deep-pencil beam and medium-large shallow surveys. In both the
0.5-2 keV and 2-10 keV bands, one of the major achievements of the XRT
survey is the improvement in the knowledge of the bright end number
counts. In the 0.5-2 keV band, at fluxes less than
$\sim$3-4$\times$10$^{-14}$ erg cm$^{-2}$ s$^{-1}$, the XRT number
counts are fully consistent within 1 $\sigma$ errors with previous
results.
At the brightest fluxes the XRT number counts are systematically lower
than the corresponding counts from the largest surveys, which should
not suffer cosmic variance as pencil beam or medium area surveys. This
systematic behavior can be due to the fact that the XRT catalog
includes only point-like sources, thus the number counts do not
include the cluster contribution (up to 20-30$\%$ at energy $<$ 2 keV
and flux $\ge$10$^{-13}$ erg cm$^{-2}$ s$^{-1}$) unlike the other
surveys. In the 2-10 keV band the number counts are consistent within
1 $\sigma$ errors with previous results at medium-deep fluxes. At the
brightest fluxes the XRT number counts are slightly lower than the high
precision Mateos et al. (2008) number counts, even if they are
marginally consistent within 1 $\sigma$ errors. This agreement, unlike
the discrepancy in the 0.5-2 keV band, is probably due to the smaller
contribution of the clusters at higher energies.

\begin{table}[ht!]
\footnotesize 
\caption{Integral number counts and sky coverage in the S and H band (see Sec. 6.2 and 6.3).}
\begin{tabular}{lcc}
\\
\hline
Flux ({\it S})   & Counts (N$>{\it S}$) & Sky coverage $\Omega_i$\\

$10^{-14}$\cgs  & deg$^{-2}$  & deg$^{2}$ \\

\hline

\multicolumn{3}{c}{0.5-2 keV}\\

\hline
     50.12    &    0.2$\pm$0.1   &   22.12  \\
      31.4  &      0.4$\pm$0.1 &   22.12\\
     19.68  &      0.9$\pm$0.2 &   22.11  \\
     12.33  &       1.9$\pm$0.3 &    22.11  \\
     7.724  &       4.3$\pm$0.4 &    22.1\\
     4.839  &       8.2$\pm$0.6 &    22.09\\
     3.032  &       18$\pm$0.9 &   22.03 \\
       1.9  &       41$\pm$1   &   21.64\\
      1.19  &       78$\pm$2   &   19.57\\
    0.7457  &       140$\pm$3  &   16.68\\
    0.4672  &       237$\pm$4  &   11.89\\
    0.2927  &       369$\pm$5  &   5.73\\
    0.1834  &       531$\pm$9  &   1.66\\
    0.1149  &       703$\pm$17 &    0.42\\
     0.072  &       969$\pm$61 &    0.047\\

\hline

\multicolumn{3}{c}{2-10 keV}\\

\hline
       54.48    &    0.3$\pm$0.1  &  22.11 \\
       32.98    &    0.8$\pm$0.2  &   22.09 \\
       19.96    &     1.8$\pm$0.3  &  22.06 \\
       12.09    &      4.9$\pm$0.5  &    21.84\\
       7.316    &     11.3$\pm$0.7  &   20.19\\
       4.429    &     29$\pm$1  &  17.32\\
       2.681    &      73$\pm$2  &   12.28\\
       1.623    &     169$\pm$4  &    5.64\\
      0.9824    &     348$\pm$9  &   1.51\\
      0.5947    &     598$\pm$22  &  0.30\\
        0.36    &     989$\pm$91  &   0.02\\

\hline	    

\end{tabular}

\end{table}

The number counts below $\sim$10 keV were previously best fitted
with broken power laws. Following Moretti et al. (2003) we parameterized the integral number counts with two power laws with
indices $\alpha_1$, which is the slope at the bright fluxes, and
$\alpha_2$, which is the slope at the faint fluxes, joining without
discontinuities at the break flux S$_0$:

\begin{equation}
N(>S)= K \frac {{(1\times 10^{-14})}^{\alpha_1} }  {S^{\alpha_1} + {S_0}^{\alpha_1-\alpha_2} S^{\alpha_2}} ~~deg^{-2}
\end{equation} 

In order to determine the parameters $\alpha_1$, $\alpha_2$ and S$_0$
we applied a maximum likelihood algorithm to the differential number
counts corrected by the sky coverage (see e.g. Crawford et al. 1970,
Murdoch et al. 1973). Although we defined the integral number
counts, the method operates on the differential counts, that is the
number of sources in each flux range which are independent of each
other, unlike the integral number counts. Moreover, using the maximum
likelihood method (L$_{max}$), the fit is not dependent on the data
binning, and therefore we can make full use of the whole data set. The
normalization K is not a parameter of the fit, but is obtained by
imposing the condition that the number of the expected sources from
the best-fit model is equal to the observed total number of sources.

Following Carrera et al. (2007), the 1$\sigma$ uncertainties for
$\alpha_1$, $\alpha_2$ and S$_0$ are estimated from range of each
parameters around the maximum which makes $\Delta$L$_{max}=$1. For
each parameter this is performed by fixing the parameter of interest
to a value close to the best fit value and varying the rest of the
parameters until a new maximum for the likelihood is found. This
procedure is repeated for several values of the parameter until this
new maximum equals L$_{max}+1$.

\begin{figure*}[h]
\begin{tabular}{cc}
\includegraphics[angle=0,height=8truecm,width=8truecm]{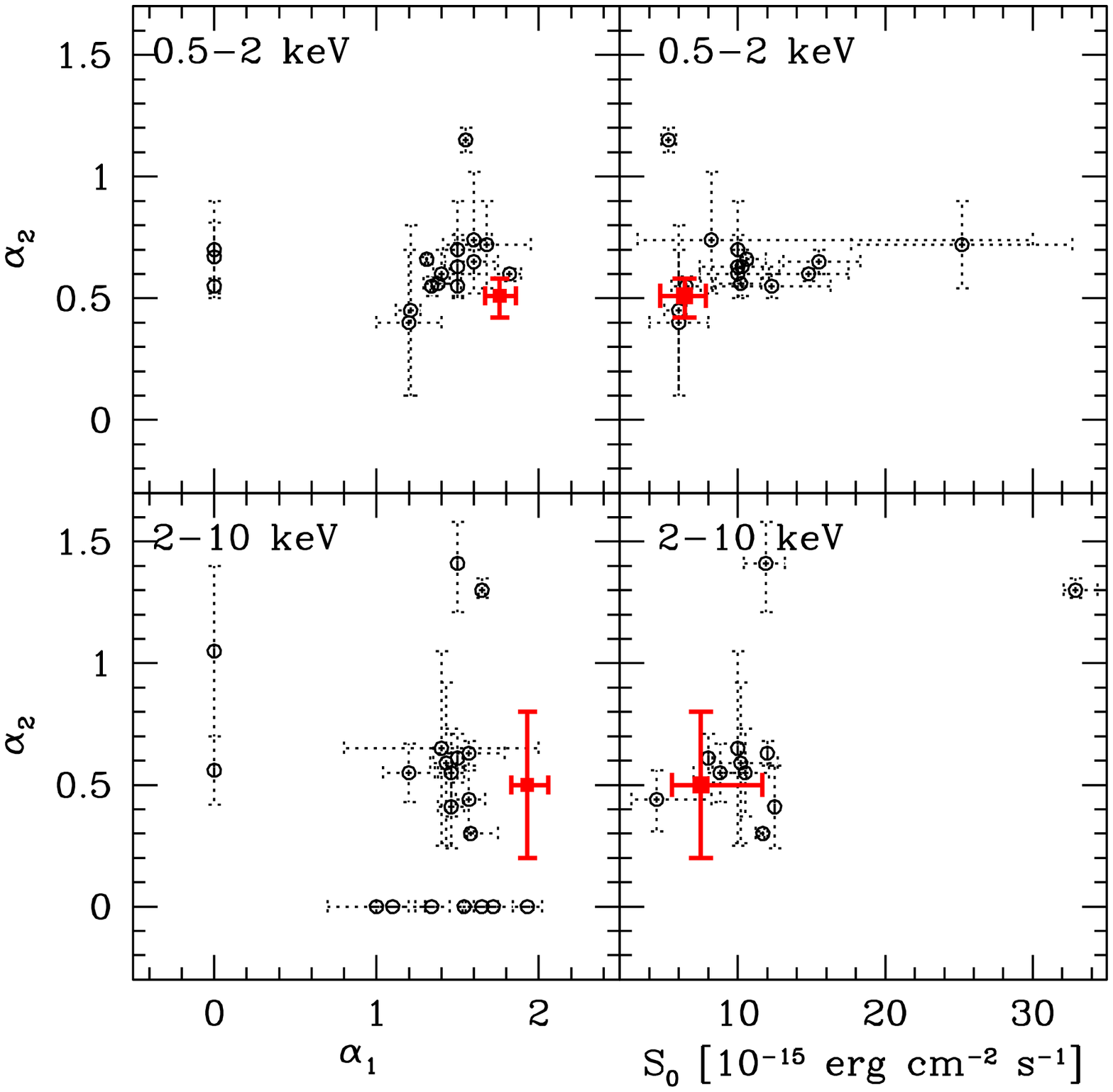}
\includegraphics[angle=0,height=8truecm,width=8truecm]{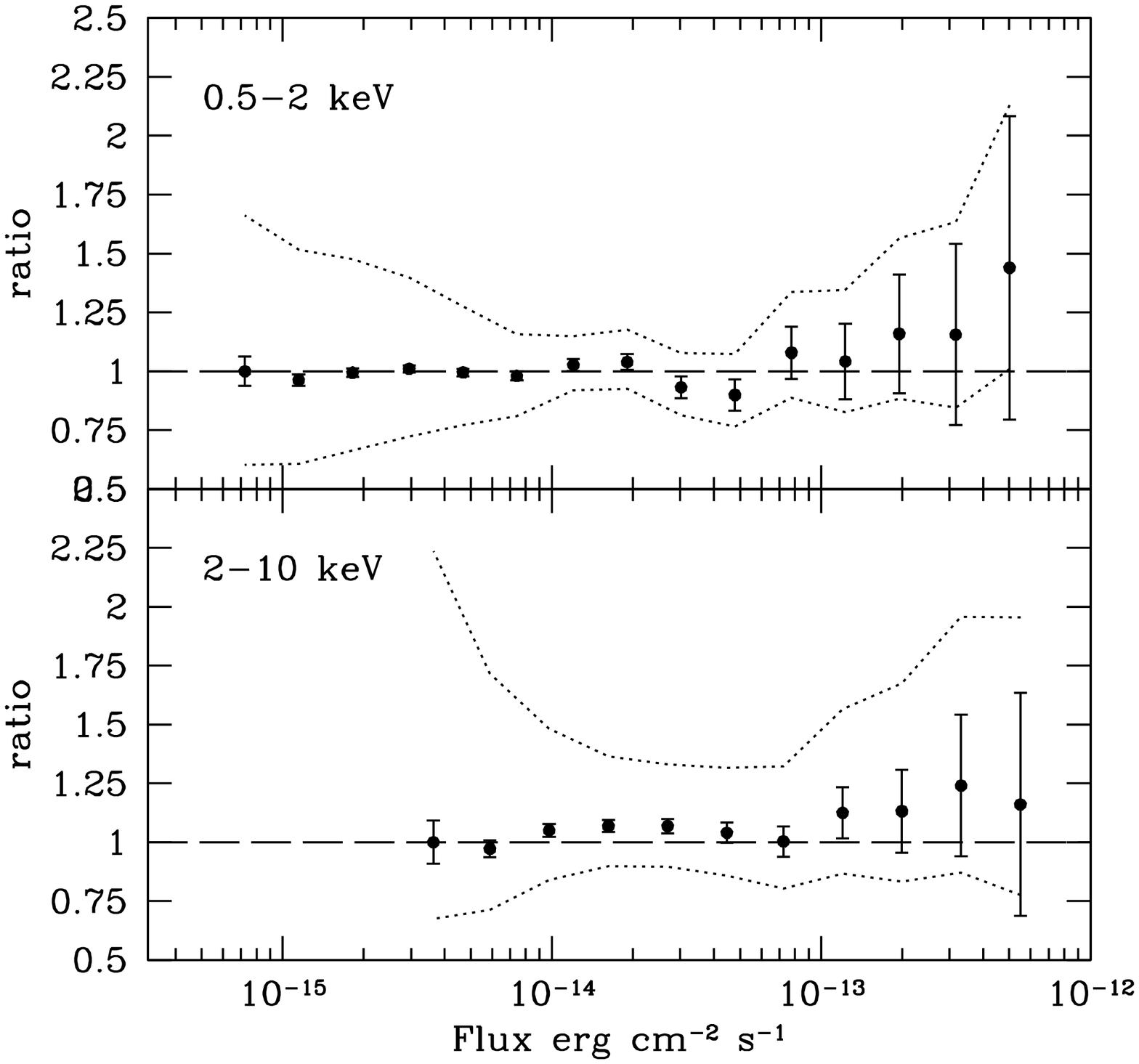}
\end{tabular}
\caption{{\it Left panel:} faint spectral index $\alpha_2$
  vs. the bright spectral index $\alpha_1$ for the 0.5-2 keV band and
  for the 2-10 keV band in the left top and left bottom panel,
  respectively, and faint spectral index $\alpha_2$ vs. the flux break
  S$_0$ for the 0.5-2 keV band and for the 2-10 keV band in the right
  top and right bottom panel, respectively (see text for the
  definition of $\alpha_1$, $\alpha_2$ and S$_0$). The red solid
  squares are our results for broken power law model, and the black
  open dots represents a compilation of data from literature: Hasinger
  et al. 1993, Giommi et al. 2000, Mushotzky et al. 2000, Page et
  al. 2000, Brandt et al. 2001, Baldi et al. 2002, Cowie et al. 2002,
  Rosati et al. 2002, Harrison et al. 2003, Moretti et al. 2003, Bauer
  et al. 2004, Kim et al. 2004, Yang et al. 2004, Hasinger et
  al. 2005, Kenter et al. 2005, Cappelluti et al. 2007, Brunner et
  al. 2008, Carrera et al. 2007, Mateos et al. 2008, Ueda et al. 2008,
  Cappelluti et al. 2009. {\it Right panel}: ratio between the binned
  integral logN-logS and the best fit model in the 0.5-2 keV band
  (upper panel) and 2-10 keV band (bottom panel). The dotted lines are
  the ratio between the binned integral logN-logS and predicted
  1-$\sigma$ uncertainty interval.}
\label{lnlsratio}

\end{figure*}

The results of the maximum likelihood fits are given in Table
\ref{lnlspar} and shown in Fig. \ref{lnlsratio}. We collected results
from previous surveys for which a logN-logS fit is available (see
Fig. \ref{lnlsratio}). We first note that the logN-logS parameters
($\alpha_1$, $\alpha_2$, and S$_0$) are not strongly constrained and
sometimes inconsistent of each other. This is probably due to the fact
that a good fit would require contemporaneous large flux coverage from
the brightest fluxes to the faintest fluxes, and in this case a more
detailed model would be necessary. Our best-fit $\alpha_2$ are
consistent at 1$\sigma$ confidence level with most of the previous
results, while our best-fit $\alpha_1$ are systematically steeper,
especially for the 2-10 keV band. Mainly for the 0.5-2 keV band, this
trend, as already noted (see text above), is probably due to the fact
that our survey does not contain clusters. The best-fit $\alpha_1$ are
steeper than the ``Euclidean slope'' of 1.5 at 1$\sigma$ confidence
level, mostly in the 2-10 keV band, probably indicating that some
amount of cosmological evolution is present (see also
Fig. \ref{lnls}).  Also our best-fit S$_0$ are consistent at 1$\sigma$
confidence level with most of the previous evaluations, further in the
0.5-2 keV band S$_0$ is better constrained and slightly lower than
previously. Note that this is not due to our higher best-fit
$\alpha_1$, in fact S$_0$ and $\alpha_1$ appear slightly positively
correlated (i.e. linear correlation coefficient of $\sim$0.47 and
$\sim$0.15 in the S and H band, respectively).

\begin{table}
\caption{LogN-logS best-fit parameters (see eq. 7).}
\begin{tabular}{lccccc}
\hline
Band$^a$ & $\alpha_1$$^b$ & $\alpha_2$$^c$ & S$_0$$^d$/10$^{-15}$ &  K$^e$ \\
keV  & & & erg cm$^{-2}$ s$^{-1}$ & deg$^{-2}$ \\
\hline
0.5-2  & 1.76$_{-0.09}^{+0.1}$  & 0.51$_{-0.09}^{+0.07}$& 6.4$_{-1.6}^{+1.4}$ & 154.9 \\
2-10  & 1.93$_{-0.10}^{+0.13}$  &  0.5$_{-0.3}^{+0.3}$ & 7.5$_{-1.9}^{+4.1}$ & 534.6\\
\hline
\end{tabular}

$^a$energy band;

$^b$power law slope for flux $\ge$S$_0$;

$^c$power law slope for flux $<$S$_0$;

$^d$flux break;

$^e$normalization factor.

\label{lnlspar}
\end{table}

Unlike the $\chi^2$ statistic, the absolute value of L$_{max}$ is not
an indicator of the goodness of fit. Then we analyzed the ratio
between the data and the best fit model (see right panel of
Fig. \ref{lnlsratio}).We did not find systematic deviations from unity
of the ratio, that would indicate that the model is not appropriate to
the data.

\subsection{X-ray spectral properties}

\begin{figure*}[h]
\begin{center}
\begin{tabular}{cc}
\includegraphics[width=7truecm]{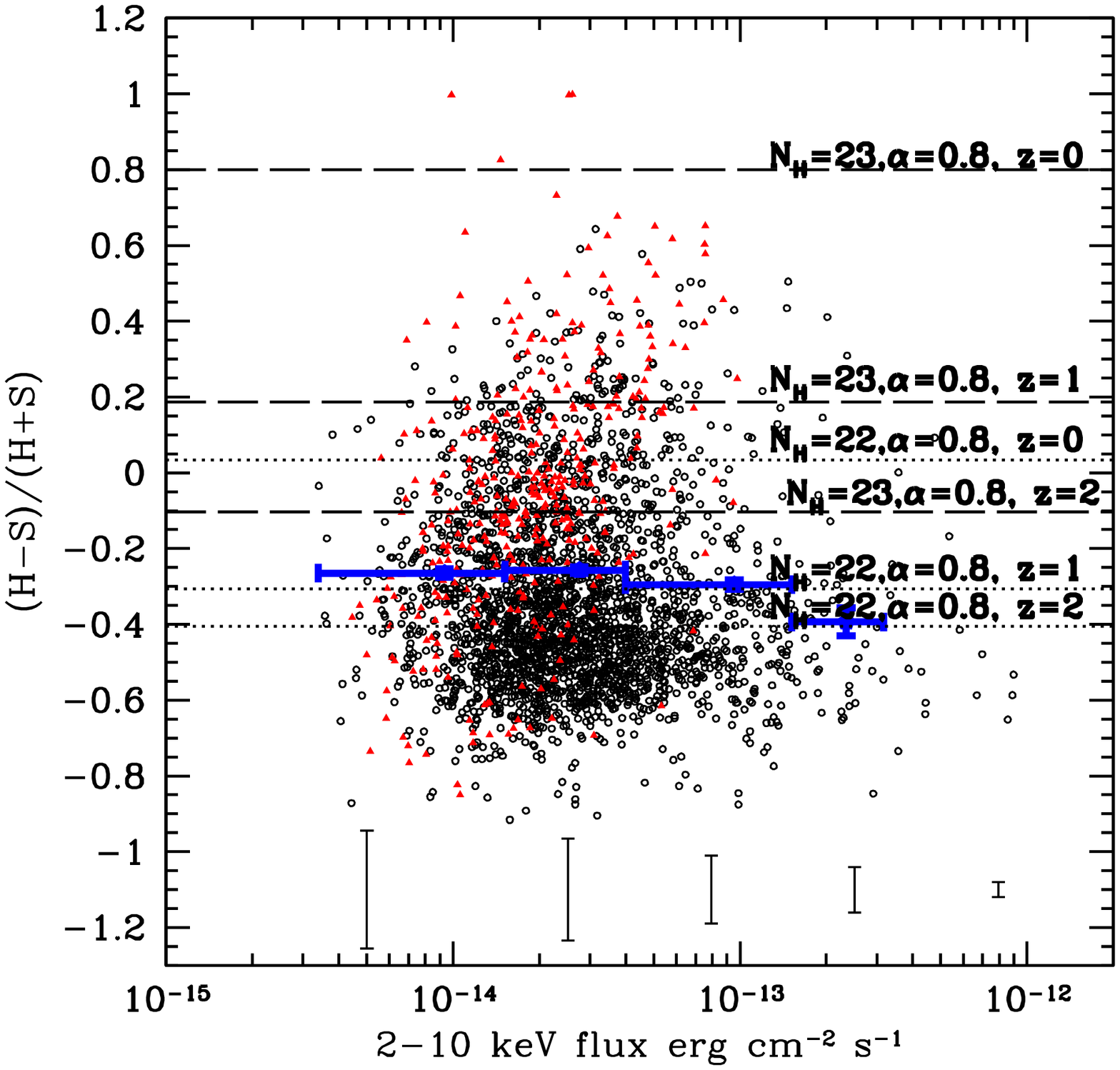}
\includegraphics[width=7truecm]{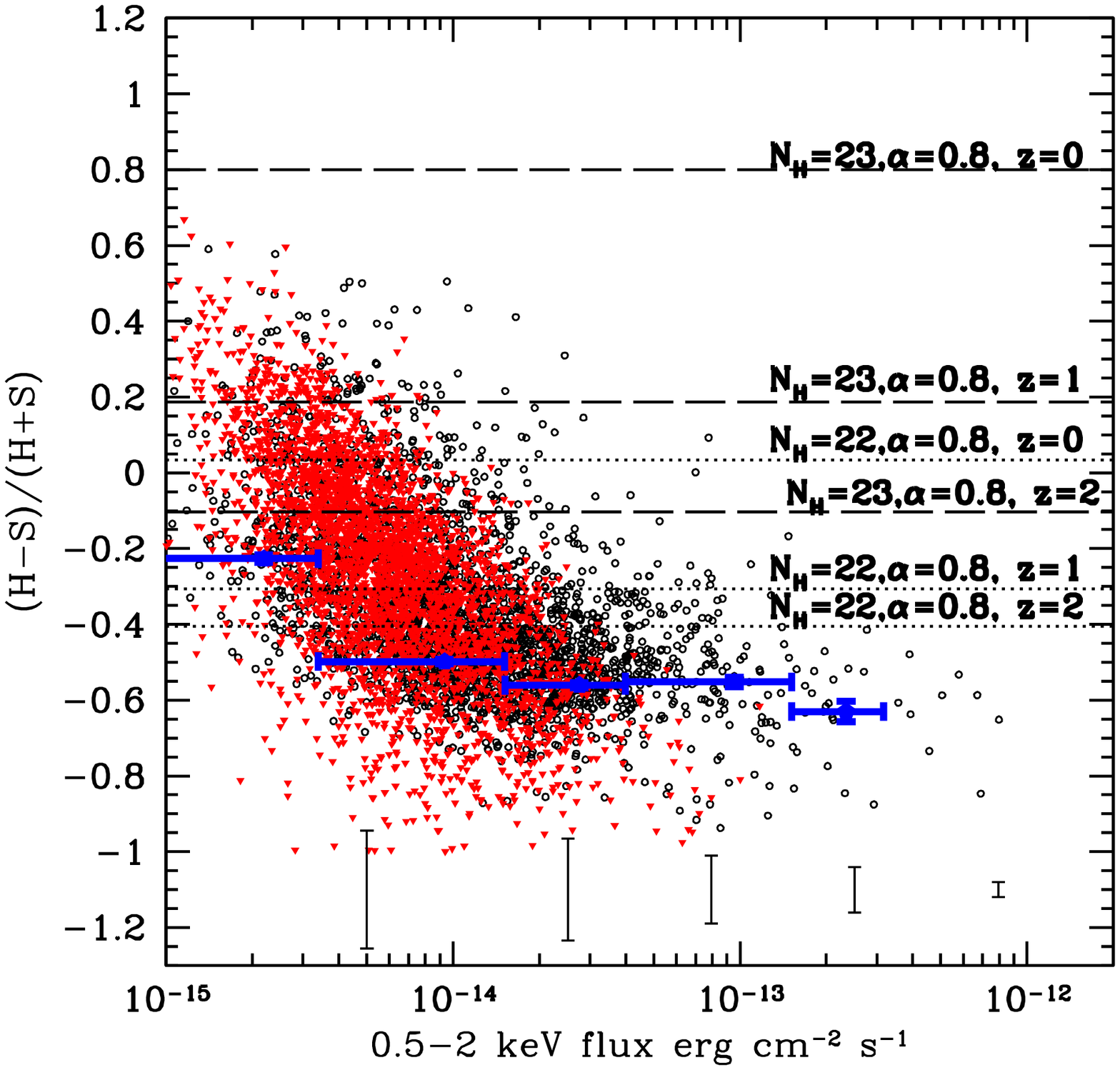}
\end{tabular}
\end{center}
\caption{{\em Left panel}: hardness ratio (see Sec. 6.4) vs. the 2-10
keV flux, for the H HGL sample. The red triangles represent lower
limits hardness ratios evaluated using 50\% S count rate upper
limits. The crosses are the mean hardness ratios in flux bins,
evaluated using the ``survival analysis'' (Kaplan \& Meier 1958;
Miller 1981, p. 74). The mean 1 $\sigma$ errors in the hardness ratio
at different fluxes are indicated by the error bars plotted at the
bottom. The dotted and dashed lines show for comparison the hardness
ratio for an absorbed power-law model of photon index $\Gamma=$1.8 and
column density 10$^{22}$ cm$^{-2}$ and 10$^{23}$ cm$^{-2}$,
respectively, at redshift decreasing from 2 to 0 (from bottom to
top). {\em Right panel}: hardness ratio (see Sec. 6.4) vs. the 0.5-2
keV flux, for the S HGL sample, here the red triangles are upper
limits hardness ratios evaluated using 50\% H count rate upper
limits. Other symbols like in right panel.}
\label{hr}
\end{figure*}

\begin{figure}[h]
\begin{center}
\includegraphics[width=7truecm]{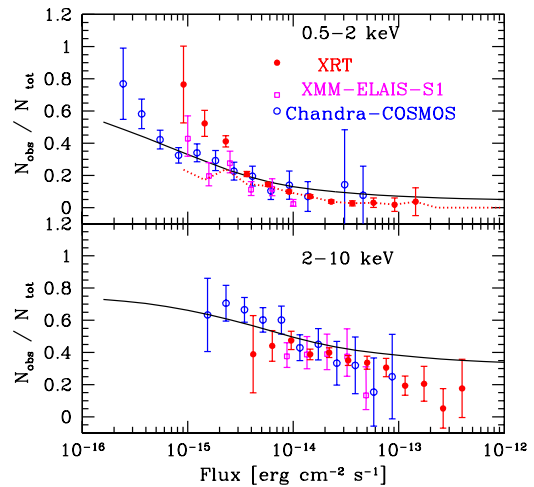}
\end{center}
\caption{Fraction of obscured sources as a function of the source flux
for three surveys: SwiftFT (red solid points), ELAIS-S1 (magenta open
points), C-COSMOS (blue open dots). The black solid line represents
the X-ray background synthesis model by Gilli et al. 2007. The red
dotted line represents the fraction of obscured sources for the
SwiftFT survey, by zeroing the flux limits in the hard band. Upper
panel: S sample. Lower panel: H sample.}
\label{frac}
\end{figure}

As a first approach, we used the hardness ratio,
HR=(c$_H$-c$_S$)/($c_H$+c$_S$) (where c$_S$ and c$_H$ are the S and H
count rates), to investigate the X-ray spectral properties of the HGL
sample. We used the ``survival analysis'' to take into account the HR
lower limits for the S sample and the HR upper limits for the H
sample. We find that the H sample shows a mean hardness ratio of
$\sim$-0.26, definitively greater than the mean HR of the S sample,
which is $\sim$-0.43. Moreover, the mean hardness ratio appears to be
anti-correlated with the flux, as in other surveys (see
e.g. HELLAS2XMM, Fiore et al. 2003, ELAIS-S1, Puccetti et al. 2006),
and in the common flux range the mean HR of the H sample is always
greater than the mean hardness ratio of the S sample. Probably this is
due to at least two reasons: 1) the contribution of non-AGN sources
with very soft X-ray colors decreases as we move to higher energies;
2) higher energies are less biased against absorbed sources, hence we
expect more absorbed sources to be detected. We also note that the
faintest S sources (see first flux bin in right panel of
Fig. \ref{hr}) have hard X-ray colors consistent with being mildly
obscured AGN.\\ Given that on the one hand the errors on HR are great,
and on the other hand the AGN spectrum can be more complex than a
simple absorbed power law model (e.g. a soft X-ray extra-component
could mimic a lower than real column density), we can roughly evaluate
the fraction of obscured sources separating them from the unobscured
sources by a threshold value of HR=-0.24, which corresponds to a
power-law model absorbed by log N$_H$$>$21.5, 22.2, 22.7 at z$=$0, 1,
2, respectively (see Hasinger et al. 2003). To take into account
sources with only count rate upper limits, we assigned each
source a count rate, that is the mean of 10000 random values, drawn
from a Gaussian distribution with mean equal to the measured count
rate and $\sigma$ equal to the count rate error or drawn from a
uniform distribution from zero to the count rate upper limit value at
50\% confidence level. We find that the fraction of obscured
sources was $\sim$37\% and $\sim$15\% for the H and S sample,
respectively. We also evaluated the fraction of obscured sources
in bin of flux (see Fig. \ref{frac}). The fraction of obscured sources
as a function of the flux is consistent within a few \% with the
results from two other surveys C-COSMOS (Elvis et al. 2009, Puccetti
et al. 2009) and ELAIS-S1 (Puccetti et al. 2006), except for the S
band, for which at flux $\le$3$\times10^{-15}$ erg cm$^{-2}$ s$^{-1}$
the fraction of obscured SwifFT sources is systematically greater than
that of the other two survey. This is probably due to the great number
($\sim57\%$) of S sources with conservative H upper limits near the
survey flux limit, because the S flux limits are deeper than the H
flux limits, this effect has an impact in the S band mainly, where a
lot of faint sources are not detected in the H band, due to the higher
H flux limit. This hypothesis is supported by the fact that the
fraction of the obscured C-COSMOS and HELLAS2XMM sources is greater
than the fraction of the obscured S SwiftFT sources, evaluated by
zeroing the H upper limits (red dotted line in the upper panel of
Fig. \ref{frac}).\\ To check whether the theoretical models show a
rough agreement with the data, we compared the fraction of the
obscured sources, defined by the hardness ratio method, as a function
of flux, with those predicted by the X-ray background synthesis model
by Gilli et al. 2007. These latter were determined by the POMPA
COUNTS\footnote{http://www.bo.astro.it/~gilli/counts.html} tool, using
a redshift range of 0-3, a column density range of 10$^{20}$-10$^{24}$
cm$^{-2}$ and a column density of 10$^{21}$ cm$^{-2}$ to distinguish
obscured from unobscured sources. The data are generally consistent
with the model predictions. The greatest discrepancy between data and
model is find in the S band, near the flux limit of each surveys (i.e
$\le$3$\times10^{-15}$ erg cm$^{-2}$ s$^{-1}$ for SwiftFT, and
$\le$4$\times10^{-16}$ erg cm$^{-2}$ s$^{-1}$ for C-COSMOS), where the
data are systematically greater by $\ge$1$\sigma$ than the
model. This, as noticed above, is probably due to the great number of
the S sources with conservative H upper limits near the survey S flux
limit.

\section{Conclusion}

We analyzed 374 {\it Swift}-XRT fields, 373 of which are
Gamma Ray Burst fields, with exposure times ranging from 10 ks to over
1 Ms. Thanks to the long exposure time of the Gamma Ray Burst fields,
the spatial isotropy of the Gamma Ray Bursts, the low XRT background,
and the nearly constant XRT PSF and vignetting, the SwiftFT can be
considered the ideal survey of serendipitous sources without bias
towards known targets, with uniform flux coverage, deep flux limit,
and large area.

Our main findings are:

\begin{itemize}

\item We produced a catalogue including the main X-ray characteristics
of the serendipitous sources in the SwiftFT. We analyzed three energy
bands S (0.3-3 keV), H (2-10 keV) and F (0.3-10 keV). We detect 9387
distinct point-like serendipitous sources, 7071 of which are at high
Galactic-latitude (i.e. $\mid$b$\mid\ge$20 deg.), with a detection
significance level $\le$2$\times$10$^{-5}$ in at least one of the
three analyzed bands, at flux limits of 7.2$\times10^{-16}$erg
cm$^{-2}$ s$^{-1}$ ($\sim$4.8$\times10^{-15}$ erg cm$^{-2}$ s$^{-1}$
at 50\% completeness), 3.4$\times10^{-15}$ ($\sim$2.6$\times10^{-14}$
erg cm$^{-2}$ s$^{-1}$ at 50\% completeness), 1.7$\times10^{-15}$ erg
cm$^{-2}$ s$^{-1}$ in the S, H, and F band, respectively. 90\% of the
sources have positional error less than 5 arcs, 68\% less than 4
arcs.

\item The large number of sources and the wide flux coverage allowed us to evaluate the X-ray number counts of the high Galactic
sample in the 0.5-2 keV and 2-10 keV bands with high statistical
significance in a large flux interval. The XRT number counts are in
agreement at 1$\sigma$ confidence level with previous surveys at faint
fluxes, and increase the knowledge of poorly known bright end of the
X-ray number counts. The integral logN-logS is well fitted (see
Fig. \ref{lnlsratio})  with a broken power law with indices $\alpha_1$ and $\alpha_2$
for the bright and faint parts, and break flux S$_0$ (see eq. 7).
Using a maximum likelihood, we find for the 0.5-2 keV band
$\alpha_1=1.76_{-0.09}^{+0.1}$, $\alpha_2=0.51_{-0.09}^{+0.07}$,
S$_0=6.4_{-1.6}^{+1.4}$ 10$^{-15}$ erg cm$^{-2}$ s$^{-1}$, and for the
2-10 keV $\alpha_1=1.93_{-0.10}^{+0.13}$, $\alpha_2=0.5_{-0.3}^{+0.3}$
and S$_0=7.5_{-1.9}^{+4.1}$ 10$^{-15}$ erg cm$^{-2}$ s$^{-1}$.

Compared to results from previous surveys, our best-fit $\alpha_2$
values are consistent at 1$\sigma$ confidence level, while our
best-fit $\alpha_1$ values are systematically steeper,
especially for the 2-10 keV band. Also our best-fit S$_0$ values are
consistent at 1$\sigma$ confidence level with most of the previous
evaluations, further in the 0.5-2 keV band S$_0$ is better constrained
and slightly lower than previously. Mainly for the 0.5-2 keV
band, the steeper value of $\alpha_1$ is probably due to the
fact that our survey does not contain clusters, unlike the other
surveys, which contribute up to 20-30$\%$ at energy $<$ 2 keV and flux
$\ge$10$^{-13}$ erg cm$^{-2}$ s$^{-1}$. The greater $\alpha_1$ and the
lower S$_0$ are not due to an intrinsic anticorrelation of the two
parameters in the model.  We note a great dispersion of the previous
logN-logS parameters (see Fig. \ref{lnlsratio}).

\item We used the X-ray colors to roughly study the obscured
sources in the HGL sample. From this analysis we find that many
sources show X-ray colors consistent with being moderately obscured
active galactic nuclei, $\sim$37\% and $\sim$15\% of the H and S
sample, respectively. The fraction of obscured sources is increasing
at low X-ray fluxes and at high energies, consistent with the results
of other surveys (see e.g. ELAIS-S1, Puccetti et al. 2006, C-COSMOS
Elvis et al. 2009). The fraction of obscured sources,
defined by the hardness ratio method, is roughly consistent with
those predicted by the X-ray background synthesis model by Gilli et
al. 2007, using rest frame hydrogen column density to define obscured
sources. A more detailed comparison between model and data, will be
possible using the sub-sample of 40\% of the high Galactic-latitude
fields, which have Sloan Sky Digital Survey coverage. For this field an
analysis of the optical counterparts is in progress.

\end{itemize}

\begin{acknowledgements}

SP acknowledges F. Fiore for the useful discussions. JPO acknowledges
the support of the STFC. We acknowledge the anonymous referee for
his comments, that helped improving the quality of the manuscript. 
\end{acknowledgements}

\end{document}